\documentclass[sigconf]{acmart}

\usepackage{multirow}
\usepackage{subfigure}
\usepackage[ruled,vlined]{algorithm2e}  
\usepackage{threeparttable}
\usepackage{enumitem}

\usepackage{booktabs}
\usepackage[normalem]{ulem}
\useunder{\uline}{\ul}{}

\AtBeginDocument{%
  \providecommand\BibTeX{{%
    \normalfont B\kern-0.5em{\scshape i\kern-0.25em b}\kern-0.8em\TeX}}}

\copyrightyear{2023} 
\acmYear{2023} 
\setcopyright{acmlicensed}\acmConference[KDD '23]{Proceedings of the 29th ACM SIGKDD Conference on Knowledge Discovery and Data Mining}{August 6--10, 2023}{Long Beach, CA, USA}
\acmBooktitle{Proceedings of the 29th ACM SIGKDD Conference on Knowledge Discovery and Data Mining (KDD '23), August 6--10, 2023, Long Beach, CA, USA}
\acmPrice{15.00}
\acmDOI{10.1145/3580305.3599453}
\acmISBN{979-8-4007-0103-0/23/08}

\begin{document}

\title{One for All: Unified Workload Prediction for Dynamic Multi-tenant Edge Cloud Platforms}


\author{Shaoyuan Huang}
\affiliation{%
  \institution{Tianjin University}
  \city{Tianjin}
  \country{China}}
\email{hsy_23@tju.edu.cn}

\author{Zheng Wang}
\affiliation{%
  \institution{Tianjin University}
  \city{Tianjin}
  \country{China}}
\email{wz_424@tju.edu.cn}

\author{Heng Zhang}
\affiliation{%
  \institution{Tianjin University}
  \city{Tianjin}
  \country{China}}
\email{hengzhang@tju.edu.cn}

\author{Xiaofei Wang}
\authornote{Corresponding author.}
\affiliation{%
  \institution{Tianjin University}
  \city{Tianjin}
  \country{China}}
\email{xiaofeiwang@tju.edu.cn}

\author{Cheng Zhang}
\affiliation{%
  \institution{Tianjin University of Finance \& Economics}
  \city{Tianjin}
  \country{China}}
\email{zhangcheng@tjufe.edu.cn}

\author{Wenyu Wang}
\affiliation{%
  \institution{Paiou Cloud Computing (Shanghai) Co., Ltd}
  \city{Shanghai}
  \country{China}}
\email{wayne@pplabs.org}

\renewcommand{\shortauthors}{Shaoyuan Huang et al.}


\begin{abstract}
Workload prediction in multi-tenant edge cloud platforms (MT-ECP) is vital for efficient application deployment and resource provisioning. However, the heterogeneous application patterns, variable infrastructure performance, and frequent deployments in MT-ECP pose significant challenges for accurate and efficient workload prediction. Clustering-based methods for dynamic MT-ECP modeling often incur excessive costs due to the need to maintain numerous data clusters and models, which leads to excessive costs. Existing end-to-end time series prediction methods are challenging to provide consistent prediction performance in dynamic MT-ECP. In this paper, we propose an end-to-end framework with global pooling and static content awareness, DynEformer~\footnote{The code is published at \url{https://github.com/hsy23/KDD23_DynEformer}}, to provide a unified workload prediction scheme for dynamic MT-ECP. Meticulously designed global pooling and information merging mechanisms can effectively identify and utilize global application patterns to drive local workload predictions. The integration of static content-aware mechanisms enhances model robustness in real-world scenarios. Through experiments on five real-world datasets, DynEformer achieved state-of-the-art in the dynamic scene of MT-ECP and provided a unified end-to-end prediction scheme for MT-ECP.

\end{abstract}





\maketitle

\section{Introduction}

The potential of cloud services has expanded with cloud computing architecture development. As a practical instance of the edge cloud computing paradigm, the multi-tenant edge cloud platforms (MT-ECP) for service network infrastructure providers, content providers (CPs), and network users have shown great commercial value. The core advantage of MT-ECP is integrating miscellaneous and heterogeneous idle computing resources (such as bandwidth, CPU, memory, etc.) in the network. Unified resource integration in MT-ECP enables flexible deployment of application services, providing users with low-latency, highly reliable edge services.




\vspace*{-0.3cm}

\begin{figure}[htbp]
    \setlength{\belowcaptionskip}{-0.3cm}
    \centering
    \includegraphics[width=8cm]{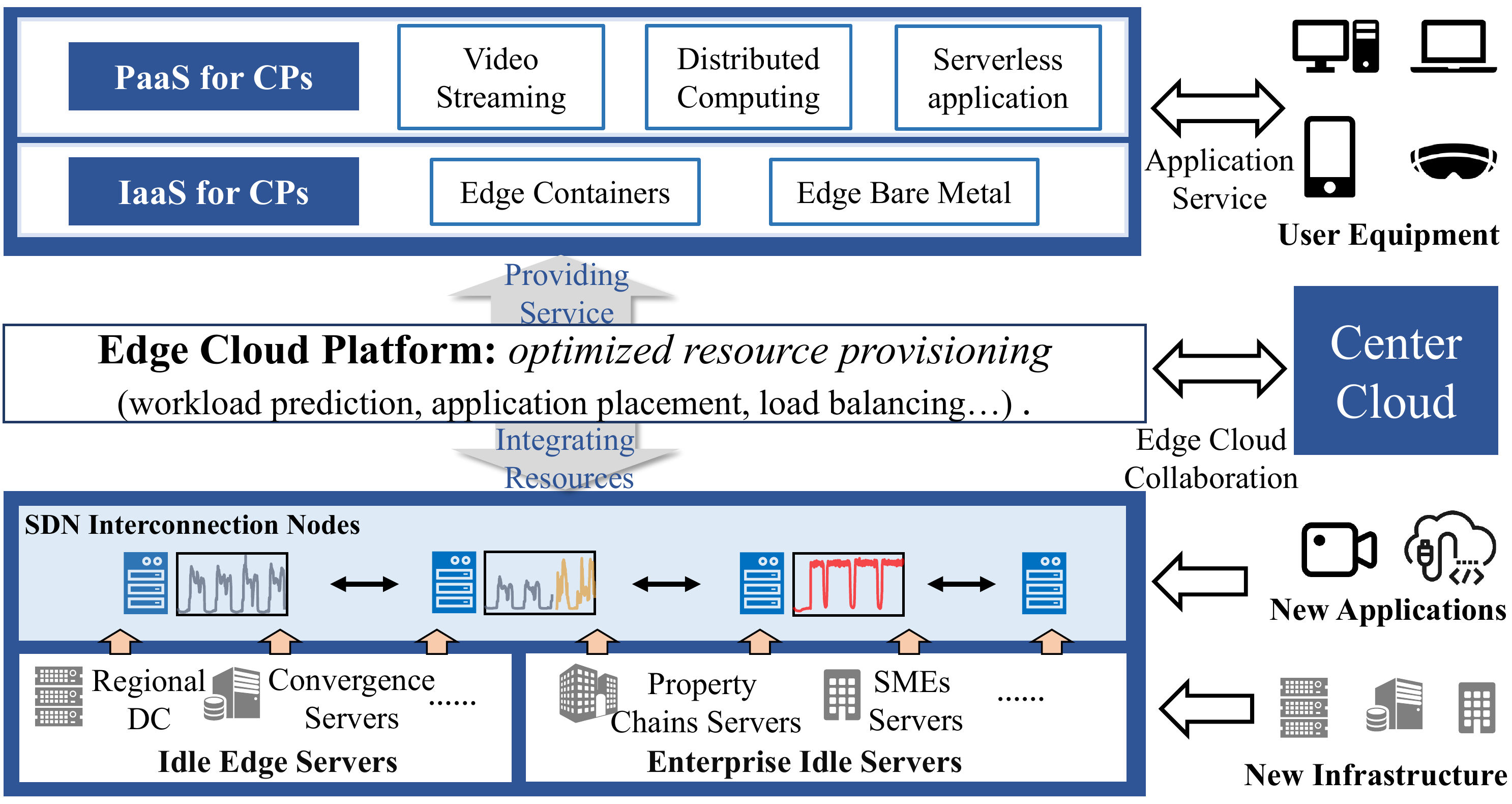}
    \caption{The multi-tenant edge cloud platform.}
    \label{fig:edge-cloud platform}
\end{figure}


As illustrated in Fig. \ref{fig:edge-cloud platform}, the role of MT-ECP is to enable the deployments of applications/services on resources. Solving this problem completely and optimally is a challenging task that requires a mix of technologies to enable attain the perception and controllability of the applications’ performance in MT-ECP~\cite{10.1145M}. 





As a necessary element of application performance perception, understanding the workload variations are greatly beneficial to the MT-ECP to solve the problems of resource planning and capacity provisioning \cite{10.1145M}. With the accurate perception of these performance indicators, application deployment and remediation tasks can be performed both proactively and effectively.

However, MT-ECP constitutes a dynamic system due to the heterogeneous application mode, differentiated infrastructure attributes, and frequent application deployments, unlike the traditional cloud service. Workload prediction, from Markov models \cite{Mumolo2017ErgodicHM} and moving averages \cite{8487450} to neural networks \cite{10.14778/3489496.3489503} and complex hybrid models \cite{yu2018improving, arbat2022wasserstein, 8457781}, has grown accurate and efficient. Although these models effectively predict workloads in stable, static deployment systems, they struggle in dynamic systems like MT-ECP.

In this paper, we focus on the workload prediction for dynamic MT-ECP. Specifically, we summarize the system dynamics of MT-ECP that lead to variations in application workloads into the following three behaviors, and predict the workload under arbitrary behaviors with a unified prediction framework.







\begin{figure}[t]
    \setlength{\belowcaptionskip}{-0.5cm}
    \centering
    \includegraphics[width=8cm]{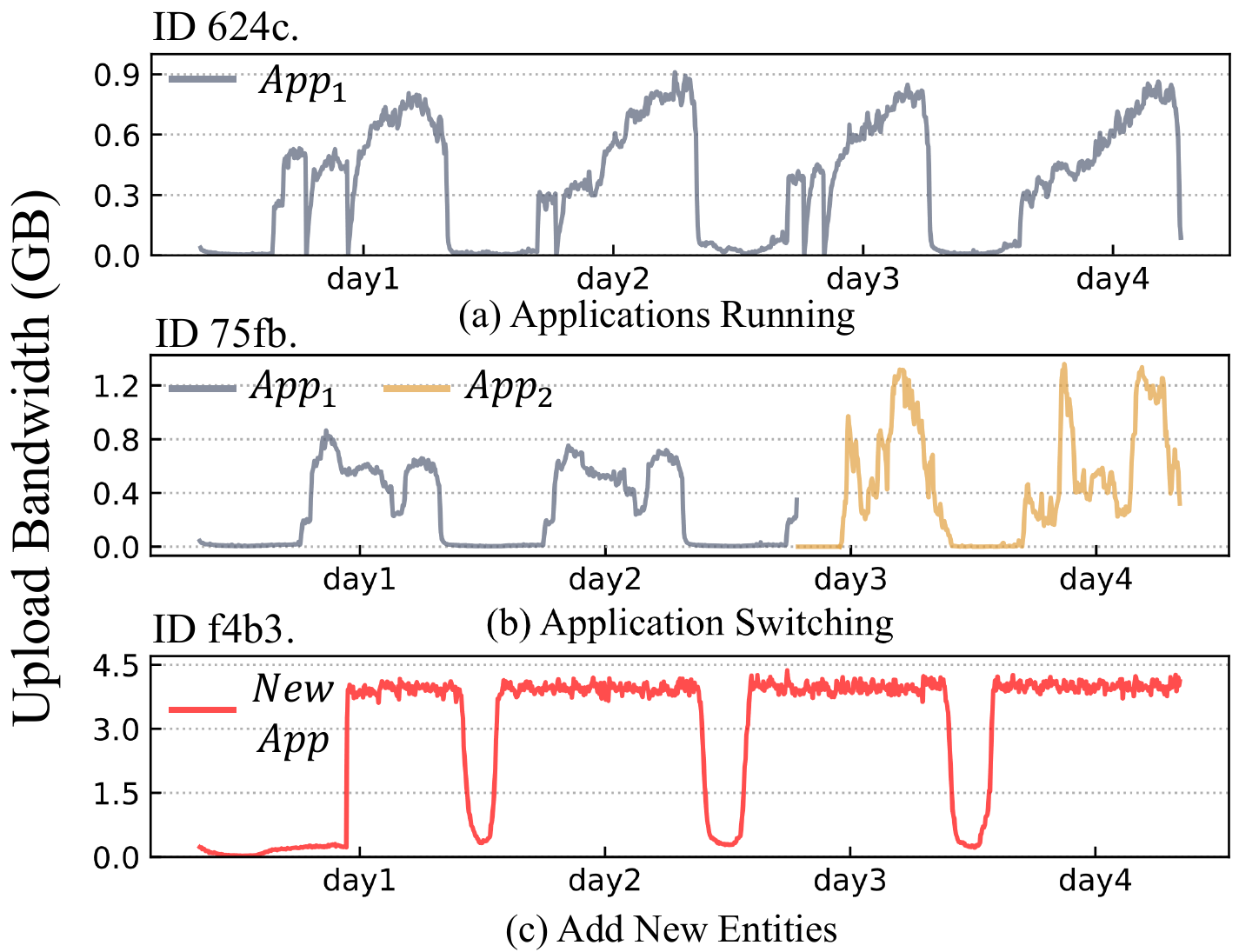}
    \caption{Workloads under dynamic MT-ECP behaviors.}
    \label{fig:workloads}
\end{figure}

\textbf{Behavior 1: Apps run on heterogeneous infrastructure:} As shown in Fig. \ref{fig:workloads}(a), the typical behavior of applications running in MT-ECP exhibits periodic workload fluctuations, guiding predictor construction. However, the MT-ECP's heterogeneous environments make the problem exceptionally complex. \textbf{Challenge 1:} The key challenge arises from the heterogeneity in applications, with varying constraints and user demand patterns, and infrastructure, with diverse hardware configurations and geographic distributions. The heterogeneity of the two levels is coupled with each other, leading to a wide variety of workload patterns.
 




\textbf{Behavior 2: Switching different apps on an edge server:} MT-ECP frequently switches apps employing automated scheduling, as depicted in Fig. \ref{fig:workloads}(b), where device 75fb.'s app shifts from $App_1$ to $App_2$ on the third day. Due to agile deployment techniques, the switching process is usually fast and non-disruptive. \textbf{Challenge 2:} New apps' workloads may vary or clash with historical patterns, exemplified by the daily peak shift from noon to night in Fig. \ref{fig:workloads}(b). In addition, new application workload data is limited but indicative of future patterns, which requires the predictor to be generalized to adjust the focus on the data. Existing work, such as research on concept drifts \cite{8496795, 9488816, SUN2021107625, 10.1145/3534678.3539348}, addresses pattern changes in dynamic circumstances. Unlike previous studies that proactively switch the model to adapt to concept drift, we want to empower the model to adapt internal weights by sensing the app switch and complete the prediction in a user-imperceptible way.




\textbf{Behavior 3: Add new entities in MT-ECP:} The new applications or new infrastructure are ubiquitous for MT-ECP. They imply unique features and few historical data. As shown in Fig. \ref{fig:workloads} (c), the workload of new applications may show a pattern that has never occurred before. \textbf{Challenge 3:} How to quickly and accurately implement workload prediction on new entities involves the cold start of the model, which is a challenge for the predictor.



To tackle the above problems, we propose DynEformer, an accurate and robust workload prediction framework for Dynamic multi-tenant Edge cloud platforms. The core of DynEformer lies in three key points: \textbf{1.} Identify the global workload patterns of the application by global pooling. \textbf{2.} Exploit the global pattern information to drive the local workload prediction under dynamic application behaviors. \textbf{3.} Improve the robustness of the model in realistic scenarios through static content awareness.



In multi-series prediction tasks, clustering-based approaches are regarded as an effective way to improve model accuracy because of their potential to exploit similarities between time series, i.e., patterns. However, existing works apply clustering to the model inputs, i.e., clustering raw inputs and training multiple models for different data classes \cite{9488816}. With numerous classes of workloads in the heterogeneous MT-ECP scenario, it is inefficient and over-costing to create and maintain multiple models simultaneously.



Therefore, we design a novel clustering alternative paradigm. We propose extracting workload's seasonal part via series decomposition and identifying global patterns by a novel global pooling method. Further, instead of creating multiple models, we design a new adaptive information merge mechanism to exploit pattern similarities/differences, adapt internal model weights to suit dynamic MT-ECP behaviors and drive workload predictions. 





In MT-ECP, static data like infrastructure configuration and location are high-semantic. Existing approaches mark inputs with static features, but effectiveness improvement is minimal as high-semantic data isn't fully understood through simple encoding. To address this, we design a static feature fusion method, allowing the model to select suitable static features for temporal workloads, supplementing static context in dynamic MT-ECP behaviors.



For historical information capture, we adopt the encoder-decoder architecture based on the transformer for effective modeling of both long- and short-term series. Our main contributions are as follows:

\begin{itemize}[leftmargin=*]
\item We propose a novel global pooling method and adaptive information merging architecture to incorporate global application patterns for the model and drive the local workload prediction under dynamic MT-ECP behaviors. This method is an effective alternative to the clustering-based approaches.
\item To adopt the cross-domain static features, a new static feature fusion method is designed that allows the model to select the most appropriate static features for the current temporal features.
\item A novel DynEformer is proposed, a global-pooling-based and static-context-aware transformer model for workload prediction under dynamic MT-ECP. DynEformer achieves a 42\% relative improvement on six benchmarks, covering five real-world datasets. In particular, it reaches a 52\% improvement under the application switching or new entity adding behaviors.
\end{itemize}

\section{Related Work}\label{rw}
In this section, we review previous approaches regarding workload analysis and prediction and encoder-decoder-based predictors.

\textbf{Workload analysis and prediction.} A large number of existing works on workload can be divided into analytical modeling and prediction of workload. The former often relies on real large-scale data (e.g., Google \cite{7562213, LIU201735}, Facebook \cite{6559990} and Wikimedia \cite{6881647}) to classify and describe workload through mathematical approaches (e.g., linear regression \cite{6005369}, and hidden Markov model \cite{Mumolo2017ErgodicHM}) and traditional machine learning models (e.g., k-means \cite{6983058} and ARIMA \cite{8487450}), with the aim of analytically answering application optimization, system self-management, and other high-level questions.



The implementation of workload prediction is more complex, as the workload variation modeling requires considering the characteristics of applications, infrastructure, and the interaction between them \cite{10.1145M}. Gao et al. \cite{9209730} and Yu et al. \cite{8377827} propose to cluster the workloads and build independent predictors per cluster. The latter also proposes to match the clusters for newly added workloads based on the initial workload patterns and static features. These works are based on traditional clustering methods, such as density-based and distance-based models, whose clustering results do not iterate over time, limiting models' long-term validity. In addition, these works cannot solve the problem of concept drift well.



Yang et al. \cite{9488816} incorporates RNN into VaDE and proposes a new workload clustering model and dynamically assigns prediction models to individual workloads based on the clustering results, which improves the accuracy of the model and can partially solve the problems of concept drift and unseen patterns. Jayakumar et al. \cite{jayakumar2020self} proposes a generic workload prediction framework that uses LSTM models and automatically optimizes its internal parameters for individual workloads through model retraining to achieve high accuracy. The main contributions of the above works lie in the pre-prediction preparation or the model-building mechanism, which does not design a unified prediction model. They still need to train and maintain multiple predictors for use, which increases the cost.




\textbf{Encoder-decoder-based predictors.} Recurrent neural networ-ks (RNNs) have been the mainstay of deep time series models until researchers started to explore introducing the transformer's design into time series prediction \cite{wu2020deep}. Due to the self-attention mechanism, transformer-based models can model both long- and short-term time series. Besides, the encoder-decoder architecture is well suited to the time series data structure.

Wu et al. \cite{wu2021autoformer} introduces time decomposition techniques and auto-correlation mechanisms into the transformer to improve its efficiency and accuracy. Lim et al. \cite{LIM20211748} proposes incorporating multiple features (including temporal and static features) from cross domains in realistic scenes into a self-attention encoder-decoder-based architecture through feature selection and recurrent layers. 


In the real MT-ECP, workloads vary with dynamic behaviors such as application switching and new application/infrastru-cture accessions. These behaviors can lead to problems such as multi-patterns, concept drift, and cold start. The works optimized for model efficiency and accuracy of idealized problems are failed to provide a unifed solution, leading to unpredictable failures, which are unacceptable in real business service platforms.


\section{Notations and Problem Definition}\label{npd}


The goal of workload prediction is to predict the future workload series, which can be defined as the bandwidth, CPU and other hardware usage or the job arrival rates (JARs) under a series of intervals. Due to the introduction of static and inferrable features such as date, the problem can be defined as a multivariate multi-series prediction problem. 

Under the rolling forecasting setting with a fixed size window, multiple workload series for different applications on infrastructures are processed as historical inputs: 
$$\mathbf{X} = \left\{\mathcal{X}_t^l = [x_t^l,...,x_{t+T-1}^l]\  \Big| \ t\leq \mathcal{T}-T+1, l \in [1, N], x_i^l \in \mathbb{R}^{d_t} \right\}$$where $\mathcal{T}$ is the last time index observable in the history, $N$ is the number of workload series and $d_t$ is the dynamic feature dimension ($d_t > 1$). Static features are presented by $\mathbf{S} = \{S^l \mid l \in [1, N], S^l\in \mathbb{R}^{d_s}\} $ and for a set of specific inputs, the prediction process can be expressed as,
$$\hat{\mathcal{Y}_t^l} = f(\mathcal{X}_t^l, S^l)$$where $\hat{\mathcal{Y}_t^l} = \{\hat{y}_t^l,...,\hat{y}_{t+L-1}^l\}$ is the predicted workload series, $L$ is the prediction length, and $f$ is a predictive model.


\section{Methodology}\label{section:DynEformer}
In this section, we present the \textbf{global pooling} method and the architecture of the \textbf{DynEformer}. As aforementioned, the global pool is built to identify the global workload patterns of applications.

Unlike previous clustering-based prediction methods, we do not train multiple models for different clusters but rather use them as the global pattern supplement for the local workload prediction. To this end, we design the \textbf{Global-Pool-merge layer (GP layer)} and implement it as a companion block of the model's encode layer to iteratively incorporate global pattern information and drive the local workload prediction. To further enhance the adaptability of the model to dynamic behaviors, we reuse the global pool to supplement additional data and design a synchronous padding mechanism that allows the model to read the changed workload data "in advance."

Besides the above designs, the \textbf{Static-context-Aware layer (SA layer)} is also incorporated into DynEformer further to enhance its prediction robustness for dynamic MT-ECP behaviors.





\begin{figure}[hbp]
    \centering
    \includegraphics[width=\columnwidth]{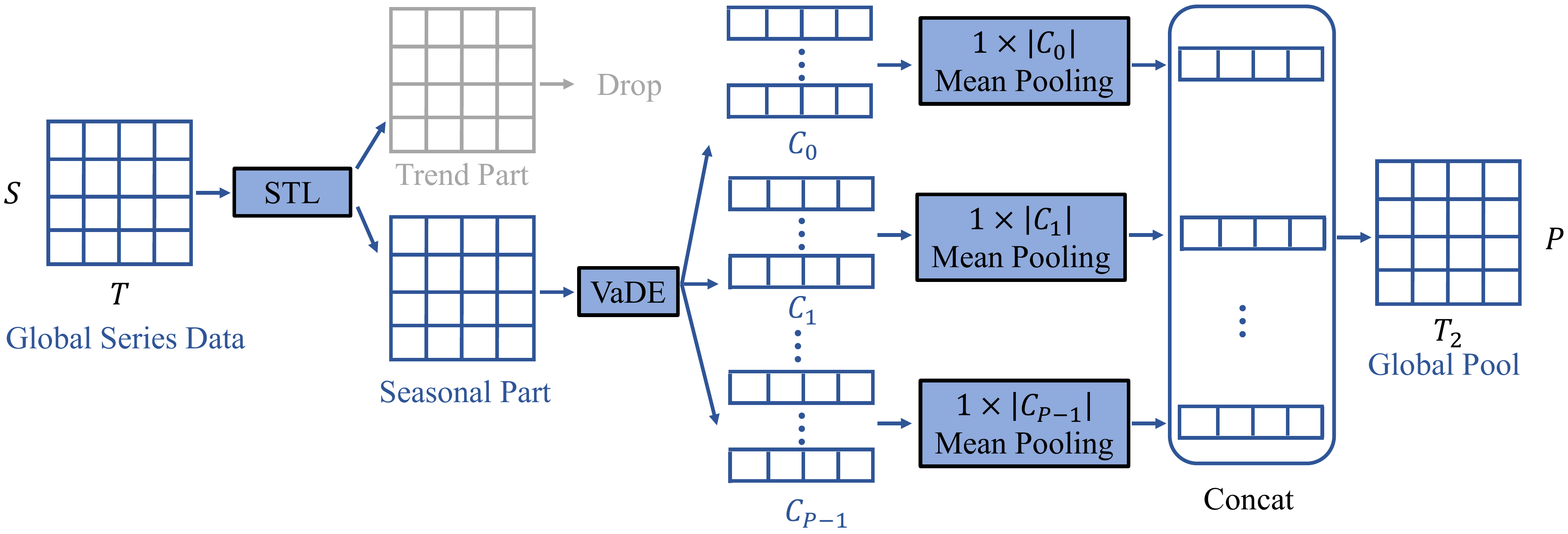}
    \caption{The process of building the global pool.}
    \label{fig:gp}
\end{figure}

\begin{figure*}[h]
    \centering
    \includegraphics[width=\textwidth]{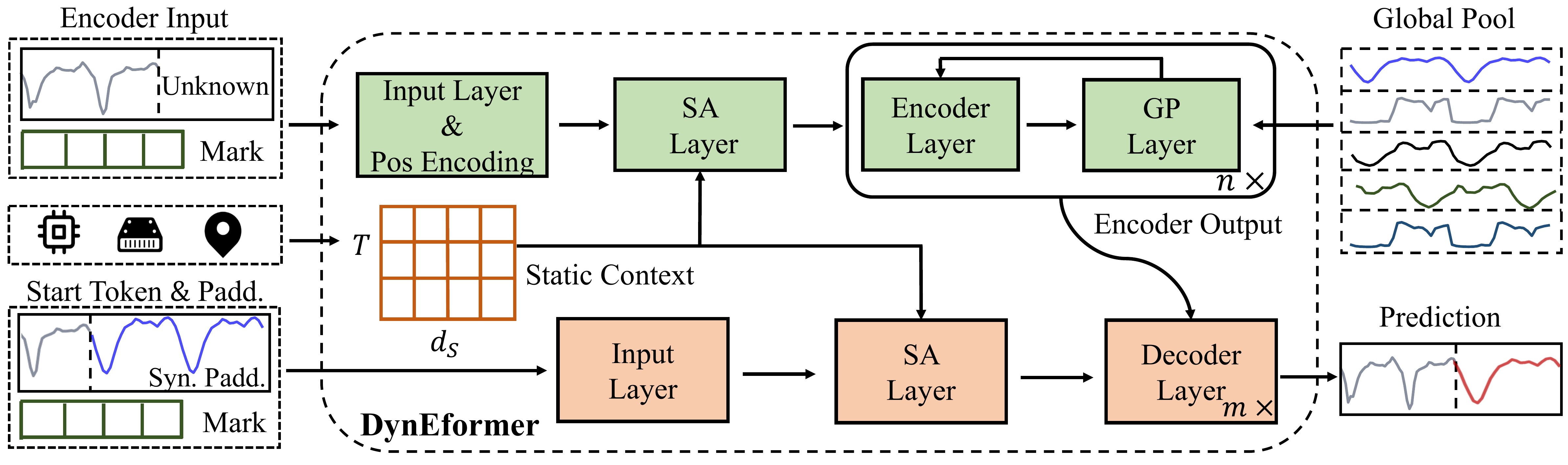}
    \caption{Framework overview of DynEformer.}
    \label{fig:framework architecture}
\end{figure*}

\subsection{Global Pooling}

\subsubsection{\textbf{Time-series decomposition}} 

Discovering the key components of a large amount of multi-pattern workloads data is necessary before building the global pool, as the model already learns the original data. We introduce decomposition techniques from traditional temporal modeling to uncover valuable information in the MT-ECP workloads. Specifically, we apply the Seasonal and Trend decomposition using Loess (STL) to divide the workload series into more fine-grained components. Given the periodicity of the workload, the decomposition procedure can be expressed as follows:
\begin{equation}
\mathcal{X}^{S}, \mathcal{X}^{T}, \mathcal{X}^{R} \leftarrow \mathbf{STL}\left(\mathcal{X}\right)
\end{equation}where $\mathcal{X}$ is the workload series, and $\mathcal{X}^{S}, \mathcal{X}^{T}, \mathcal{X}^{R}$ represent the seasonal component, trend component, and residual component of $\mathcal{X}$, separately.


The residual component $x_t^R$ is not considered due to the low information. For the other components, we distinguish that in the MT-ECP workloads, the trend component $x^T$ is highly random, and the pattern information is low. In a long time, all workloads have only three categories: smooth, increasing, or decreasing, but in a shorter time, they have a large number of random changes making it difficult to categorize. For the seasonal component $x^S$, we find that the application type mainly reflects it. The seasonal component has distinct features that allow for effective classification when setting the daily period. Therefore, we extract the seasonal component of the workload as the source to build the global pool.

\subsubsection{\textbf{Clustering with VaDE}}

In building the global pool, we cluster the seasonal component of workloads (but do not build multiple models based on the clustering results). To achieve this, we use VaDE \cite{ijcai2017p273}, a clustering method that differs from traditional distance-based or density-based clustering methods and consists of encode, decoder and Gaussian Mixture Model (GMM).
Before the seasonal component is input to the VaDE, we do the same sliding window processing with a window size of $T_2$. Details of VaDE are shown as follows:
\begin{equation}
\begin{split}
Z_t&=\mathbf{Encoder}\left(\mathcal{X}_t^{S} \right) \\
C&=\mathbf{GMM}\left(Z_t \right) \\
\hat{\mathcal{X}}_t^{S}&=\mathbf{Decoder}\left(Z_t \right)
\end{split}
\end{equation}where $\mathbf{Encoder}$ and $\mathbf{Decoder}$ are blocks consisting of several different full-connected (fc) layers, $Z_t$ is the encoding vector of the workload series, and $C$ is the clustering result. $\mathbf{GMM}$ assumes a Mixture of Gaussian generates the $Z_t$ decided by the latent variable $C$ and can infer the probability of $C$.


There are two advantages of VaDE: 1. Longer time series in clustering means higher feature dimensionality. VaDE can better handle high-dimensional data and capture correlations between features. 2. When new data arrives, VaDE can update the parameters based on the pre-trained model without retraining every time, which can better support model updates over a long period.

\subsubsection{\textbf{Building the global pool}}

The entire process of global pool building is shown in Fig. \ref{fig:gp}. To choose the optimal number of clusters $P$, we empirically infer the Bayesian Information Criterion (BIC) and select $P$ where the model's BIC is lower than other cases. After VaDE outputs the category of each series, we average the pooling for all the series in each category, compressing them into one representative series, after which the representative series of all categories are concated together into the global pool as follows:
\begin{equation}
\mathcal{P} = \left\{\left[\mathbf{Mean Pooling}(\mathcal{X}_t^{S, l})\ |\ \mathcal{X}_t^{S, l} \in C_i\right] \Big|\ i \in [1,2,..., P] \right\} \nonumber
\end{equation}
Note that the global pool is built entirely on the training (historical) data, and any workloads in the test (future) data will not be seen by the global pooling in the experiments.

\subsection{DynEformer}
In this section, we introduce our DynEformer. The overview of DynEformer is shown in Fig. \ref{fig:framework architecture}, which consists of two main parts: Encoder and Decoder. The inputs of DynEformer include encoder input, decoder input, static content, and global pool, and the predicted workloads are output at the end of the framework process. DynEformer is a complete end-to-end framework that provides unified workload prediction for all dynamic behaviors of MT-ECP.


\subsubsection{\textbf{Encoder}}
The encoder of the DynEformer is composed of three components: Input \& positional encoding layer, Static-context-Aware layer (SA layer), and a stack of $n$ identical GP-encoder blocks.

\textbf{Input \& positional encoding layer:} The input and positional encoding layer follow the original Transformer architecture \cite{NIPS2017_3f5ee243}. The input layer is used to map the past workload series to a vector of dimension $d_{model}$ through a fully-connected network. This step is essential for the multi-head attention mechanism. Positional encoding with sine and cosine functions is used to encode sequential information by element-wise addition of the input vector with a positional encoding vector.

\vspace{-0.3cm}
\begin{figure}[htbp]
    \centering
    \setlength{\belowcaptionskip}{-0.3cm}
    \includegraphics[width=4cm]{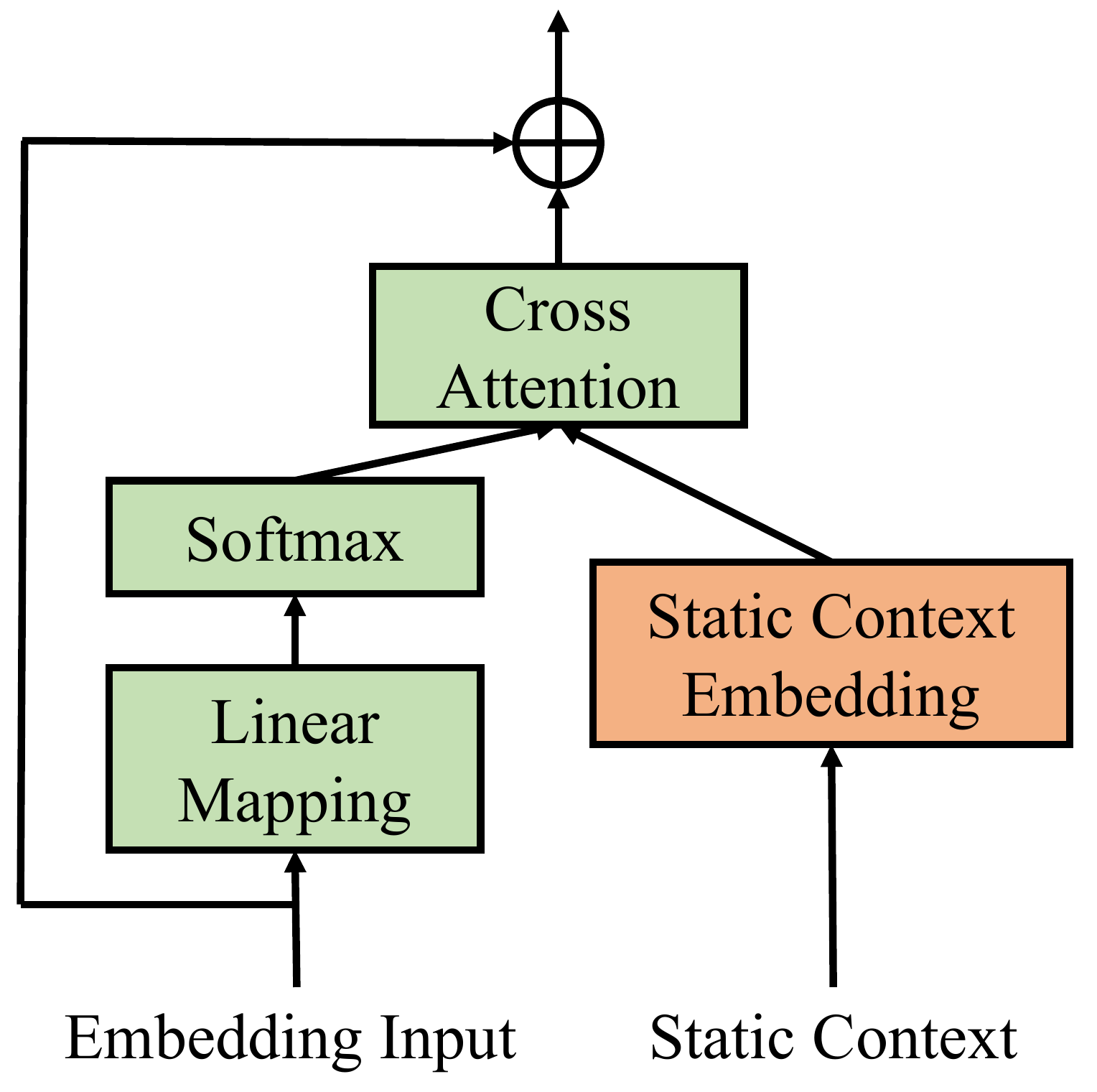}
    \caption{SA Layer.}
    \label{fig:sa}
\end{figure}


\textbf{SA Layer:} The SA layer is used to combine the static context with the varying workload series data. We introduce cross-domain data (e.g., infrastructure's maximum bandwidth, memory, hard disk and geographic location) recorded from the MT-ECP as additional static content. To learn these high-semantic data, we design the static-context-aware layer (SA layer), which uses cross-attention to select the most efficient static content to integrate based on the current series embedding. The input of the SA layer contains the workload embedding $\mathcal{V}_t$ passing through the input layer and the static context matrix $\mathbf{S}$, and its output dimensional is the same as $\mathcal{V}_t$, but with additional information about the encoding of static content. Details of SA layer are shown as follows:
\begin{equation}
\begin{split}
Q_t &= \mathbf{Linear}_1(\mathcal{V}_t) \\
\alpha_t&=\mathbf{softmax}^2(Q_t) \\
V&=\mathbf{Linear}_2(\mathbf{S}) \\
\mathcal{V}_t&=\mathbf{Norm}(\mathcal{V}_t+\mathbf{Dropout}(\alpha_t V))
\end{split}
\end{equation}where $\mathbf{softmax}^2$ denotes the softmax in the 2nd dimension of $Q_t$, i.e., weights are calculated for each time point of each series sample.

Unlike the traditional method in which static content is concatenated to the inputs as marks, the proposed SA layer acts as a filter to learn the most effective static content. Meanwhile, the SA layer can assist in learning similar infrastructure attributes, and this perception of static content is beneficial to solve the concept drift and cold start problems in dynamic MT-ECP.


\begin{figure}[htbp]
    \centering
    \includegraphics[width=7cm]{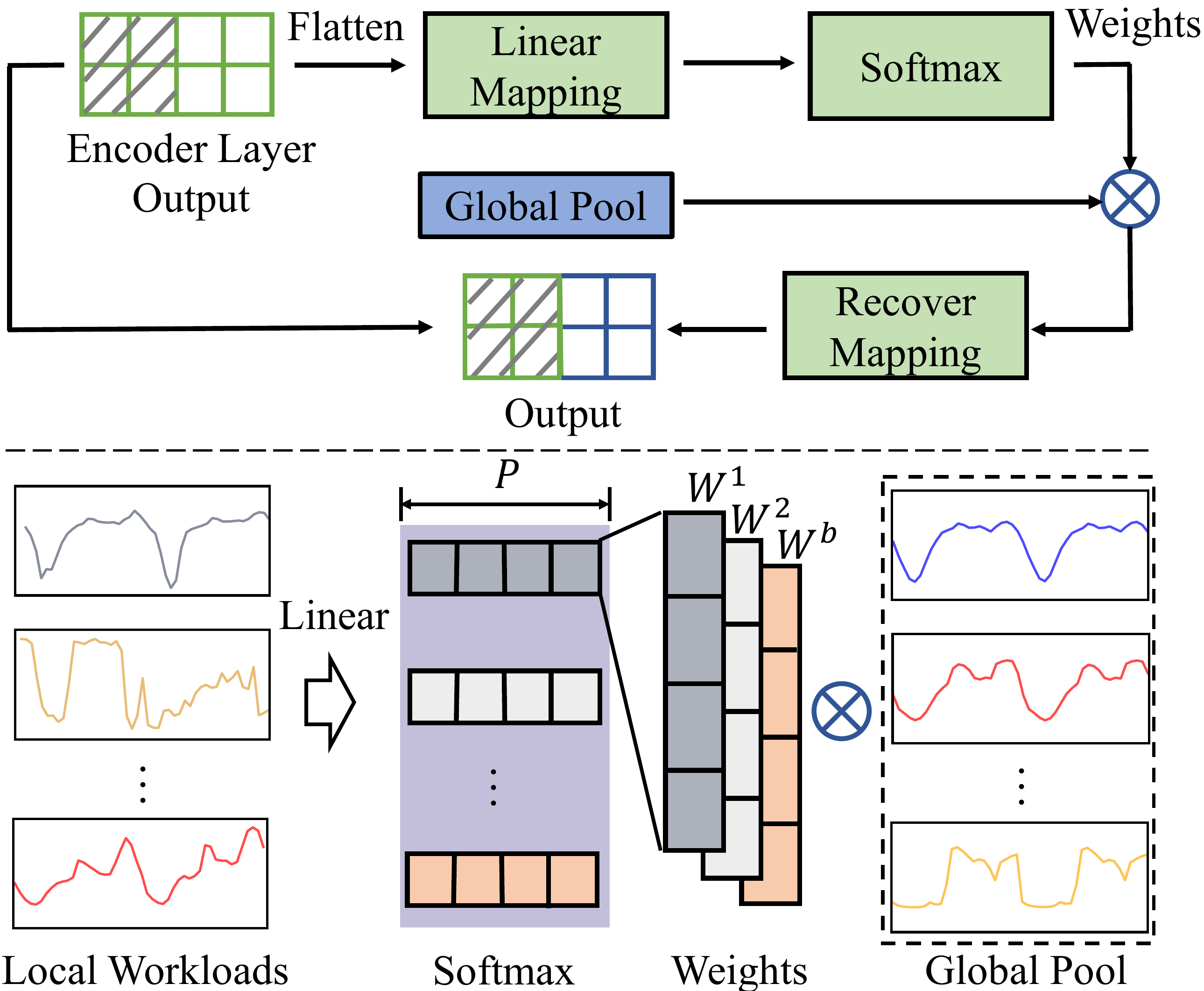}
    \caption{The Global-Pool-merge (GP) layer.}
    \label{fig:gp layer}
\end{figure}

\textbf{GP-encoder Block:} The GP-encoder block focus on driving local workload prediction by merging global pattern information. Each GP-encoder block consists of a regular encoder layer with full attention mechanism and a novel GP layer that connects the input and output of the encoder layer while continuously capturing the most relevant global application patterns from the global pool. Suppose there are $n$ encoder blocks, the calculation of the $i$-th GP-encoder block is shown as follows:
\begin{align}
E_t^{i}&=\mathbf{Encoder}\left(\mathcal{V}_t^{i-1} \right) \\
\mathcal{V}_t^{i}, W_t^{i} &=\mathbf{GP\ Layer}\left(E_t^{i}, \mathcal{P} \right)
\end{align}where when $i=1$, $\mathcal{V}_t^{i-1}$ is the output of SA layer. The GP layer takes the output of the encoder layer and the gloabl pool created in $\S$4.1 as input, the i-th GP layer can be formalized as follows:
\begin{align}
W_t^{i}&=\mathbf{softmax}^1\left(\mathbf{Linear}_1\left(\mathbf{flatten}(E_t^{i}[:,:,\frac{1}{2}d_{model}:])\right)\right) \in \mathbb{R}^{b \times P} \\
\mathcal{E}_t^{i}&=W_t^{i}(\mathcal{P}^T[:,\_,:]\leftarrow\mathcal{P})\in \mathbb{R}^{T \times b \times P} \\
\mathcal{V}_t^{i} &= \mathbf{concat}\left(E_t^{i}[:,:,:\frac{1}{2}d_{model}],\mathbf{Linear}_2(\mathcal{E}_t^{i})^T\right)\in \mathbb{R}^{b \times T \times d_{model}}
\end{align}where $E_t^{i}[:,:,\frac{1}{2}d_{model}:]$ represents the last $\frac{1}{2}$ feature dimensions of $E_t^{i}$ and $b$ is the batch size. GP layer uses this part of the features to obtain the merging weights $W_t^{i}$ and extract the global pattern information, while the remaining features will retain the temporal features. $\mathbf{flatten}$ operates on the last two dimensions of $E_t^{i}$, the results dimension is $b\times(T\times \frac{1}{2}d_{model})$, $\mathbf{softmax}^1$ denotes the softmax in the first dimension, i.e., weights are calculated for each series sample, and $[:,\_,:]$ represents the dimension raising, after which the dimension of $\mathcal{P}^T$ has changed from $T\times P$ to $T\times 1 \times P$.



GP-encoder block can significantly enhance the encoder's ability to learn global pattern information. It supersedes the conventional clustering framework and allows the DynEformer to sense the similarities and differences between global and local
patterns to adapt internal weights to suit dynamic MT-ECP behaviors without training multiple models.


\subsubsection{\textbf{Decoder}}
As shown in Fig. \ref{fig:framework architecture}, the Decoder focus on the future workload series generation. It is composed of three components: input layer, SA layer (as explained in $\S$ 4.2.1), and a stack of $m$ decoder layers. With the latent variable $\mathcal{V}_t^{n}$ from the encoder, and decoder inputs $\mathcal{X}_{t}^{de}$, the decoder can be formalized as follows:
\begin{align}
\mathcal{H}_t&=\mathbf{Linear}\left(\mathcal{X}_{t}^{de} \right) \\
H_t &=\mathbf{SA\ Layer}\left(\mathcal{H}_t, \mathbf{S} \right) \\
\mathcal{D}_t^{i}&=\mathbf{Decoder}\left(H_t, \mathcal{V}_t^{n} \right)
\end{align}

\textbf{Inputs with synchronous padding:} Decoder's inputs $\mathcal{X}_{t}^{de}$ are as follows:
\begin{align}
    \mathcal{X}_{t}^{de} = \mathbf{Concat}\left(\mathcal{X}_t[:,L_{token}:,:], \mathcal{X}_{t}^{mark}, \mathcal{X}_{t}^{gp} \right) \\
    \in \mathbb{R}^{b\times (L_{token}+L_y)\times d_t} \nonumber
\end{align}where $\mathcal{X}_t[:,L_{token},:] \in \mathbb{R}^{L_{token}\times d_t}$is the start token \cite{Zhou2021}, $\mathcal{X}_{t}^{mark} \in \mathbb{R}^{L_y\times (d_t-1)}$is the series mark e.g., date, and $\mathcal{X}_{t}^{gp} \in \mathbb{R}^{L_y\times 1}$ is the synchronous padding.

Since the global information is incorporated in the encoder output $\mathcal{V}_t^{n}$, filling the $\mathcal{X}_{t}^{de}$ with the corresponding padding can effectively improve the information capture of the decoder in the encoder-decoder attention stage. Details of the synchronous padd-ing are as follows:
$$
\mathcal{X}_{t}^{gp} = W_t^{n}\mathcal{P}
$$where $W_t^{n}$ is the global pool merging weights in $\S$4.2.1.


\subsection{Training Process and Loss Funciton}


We generate the global pool in advance by pre-building, w.r.t, the training of VaDE, and the global pool building are complete before the training of the DynEformer. The design of VaDE follows the auto-encoder mechanism, and the model is trained through unsupervised learning, i.e., the parameters of the encoder and decoder are corrected by computing the loss between $\mathcal{X}_t^{S}$ and $\hat{\mathcal{X}}_t^{S}$ for gradient backpropagation.

After we get the global pool, it will be taken as a fixed input to the DynEformer. Besides the parameters of the input layer and the multi-headed attention mechanism in encoder and decoder, the SA layer and GP layer also update their optimal parameters during the training process. Both VaDE and DynEformer use the mean squared error (MSE) loss.

\section{Experiments}\label{Exp}

\begin{table}
\caption{Workload Datasets.}
\label{datasets}
\resizebox{\columnwidth}{!}{
\centering
\begin{tabular}{llccc}
\toprule
\textbf{Workload} &\textbf{Time Range} &\textbf{Series Num$^1$} &\textbf{Interval}\\ \midrule
ECW   &2022/8-2022/9   &689 &5mins\\ \bottomrule
ECW-App Switch  &2022/8/25-/8/30   &10 &5mins\\ \bottomrule
ECW-New Infras.  &2022/8/25-/8/30   &651 &5mins\\ \bottomrule
ECW-New App  &2022/9-2022/10   &11 &5mins\\ \bottomrule
Azure  &2017/8-2017/9   &1 &1mins\\ \bottomrule
\end{tabular}}
\begin{tablenotes} 
		\item $^1$\textbf{Series Num} refers to the number of workload series (prior to sliding time window).
\end{tablenotes} 
\end{table}

\textbf{Datasets.} To comprehensively measure the proposal's effectiveness under MT-ECP workload prediction, we collect the Edge Cloud upload bandwidth Workload data (ECW) from a commercial crowd-sourced edge cloud, whose infrastructure includes heterogeneous devices in the form of self-built, recruited, and rented from users. The platform contains 5174 devices distributed throughout the country and serves 40+ typical applications. The sample workloads shown in Fig. \ref{fig:workloads} are from the actual deployments under this platform.

To validate the performance of DynEformer when application switching or new entities add to the MT-ECP, we provide some unique case data, including (1) ECW-App Switch, which are the workload series where application switching occurred during the ECW test period. (2) ECW-New Infras. are the workload series running on infrastructure that has never been present in the ECW. (3) ECW-New App are the workload series of the applications that never appeared in ECW.


In addition to temporal workloads data, ECW also provides 14 dimensions of static content data, such as maximum bandwidth, number of CPUs, location, and other infrastructure attributes. Since static content data is unavailable for all workload prediction tasks, we provide an evaluation of the effectiveness of DynEformer on such tasks. We use a public cloud workload (with JARs) dataset Azure \footnote{https://github.com/Azure/AzurePublicDataset/} \cite{10.1145/3132747.3132772} to validate the generalization of the proposal for these tasks. Datasets details are shown in Table \ref{datasets}.

Though Azure is a public cloud service dataset, it still contains some common MT-ECP characteristics and challenges, such as varying application load patterns and fluctuations \cite{arbat2022wasserstein}. By using this dataset, we showcase that DynEformer is not only capable of handling the unique challenges in MT-ECP but also performs effectively in other similar time series prediction tasks. 


\begin{table*}[htbp]
\caption{Average prediction performance for edge cloud workloads. A lower MSE or MAE indicates a better prediction.}
\label{table:main result}
\begin{threeparttable}
\resizebox{\textwidth}{!}{
\begin{tabular}{@{}cccclclcccccccc@{}}
\toprule
Models                           & \multicolumn{2}{c}{\textbf{DynEformer}}                                  & \multicolumn{2}{c}{Autoformer}                        & \multicolumn{2}{c}{Informer}                          & \multicolumn{2}{c}{Deep Trans.} & \multicolumn{2}{c}{MQRNN} & \multicolumn{2}{c}{DeepAR} & \multicolumn{2}{c}{VaRDE-L} \\ \cmidrule(l){2-3} \cmidrule(l){4-5} \cmidrule(l){6-7} \cmidrule(l){8-9} \cmidrule(l){10-11} \cmidrule(l){12-13} \cmidrule(l){14-15} 
Metrics                          & MSE                                & MAE                                & MSE                       & \multicolumn{1}{c}{MAE}   & MSE                       & \multicolumn{1}{c}{MAE}   & MSE            & MAE            & MSE         & MAE         & MSE         & MAE         & MSE         & MAE          \\ \midrule
\multicolumn{1}{c|}{ECW}         & \multicolumn{1}{l}{\textbf{0.067}} & \multicolumn{1}{l}{\textbf{0.141}} & \multicolumn{1}{l}{0.079} & 0.146                     & \multicolumn{1}{l}{0.073} & 0.147                     & 0.075          & 0.151          & 0.082       & 0.163       & 1.270        & 0.750       & 0.150        & 0.218        \\ \midrule
\multicolumn{1}{c|}{App Switch}  & \textbf{0.067}                     & \textbf{0.169}                     & \multicolumn{1}{l}{0.158} & 0.270                     & \multicolumn{1}{l}{0.099} & 0.223                     & 0.146          & 0.288          & 0.335       & 0.442       & -           & -      & 0.270           & 0.369            \\ \midrule
\multicolumn{1}{c|}{New Infras.} & \textbf{0.065}                     & \textbf{0.143}                    & 0.076                     & \multicolumn{1}{c}{0.163} & 0.070                     & \multicolumn{1}{c}{0.157} & 0.071          & 0.154          & 0.076       & 0.159       & -           & -      & 0.234           & 0.348            \\ \midrule
\multicolumn{1}{c|}{New App}     & \textbf{0.090}                      & \textbf{0.207}                     & 0.111                     & \multicolumn{1}{c}{0.229} & 0.109                     & \multicolumn{1}{c}{0.235} & 0.175           & 0.308          & 1.497       & 0.861       & -           & -      & 0.556           & 0.502         \\ \midrule  
\multicolumn{1}{c|}{Azure}       & \multicolumn{1}{l}{\textbf{0.069}} & \multicolumn{1}{l}{\textbf{0.180}} & \multicolumn{1}{l}{0.495} & 0.540                     & \multicolumn{1}{l}{0.496} & 0.521                     &0.212                &0.395                & 0.225             &0.346             & 21.248            & 3.162      & 1.124           & 0.730             \\ \bottomrule
\end{tabular}}
\begin{tablenotes} 
		\item The "-" indicates the model performance is too poor to compare, excluded to prevent evaluation misdirection.
\end{tablenotes} 
\end{threeparttable} 
\end{table*}

\subsection{Implementation Details}

The workload series in all ECW datasets are partitioned hourly, and the maximum of each hour is taken to match the time granularity of application scheduling and billing rules in the MT-ECP. The series in Azure are summed by 5 minutes to reduce the 0 values. The input workload series length $T=48$ and prediction length $L=24$ for both datasets. We split the ECW and Azure into training, validation and test set in chronological order by the ratio of 6:2:2, and test the DynEformer on other ECW datasets.


The decomposition period of STL is set to 24. The data sliding windows size $T_2$ in $\S 4.1.2$ is set to $T_2=T$. DynEformer is trained with the ADAM \cite{kingma2014adam} optimizer with an initial learning rate of $10^{-4}$. Batch size is set to 256 and the start token length $L_{token}$ is set to 12. All experiments are repeated three times, implemented in PyTorch \cite{paszke2019pytorch} and conducted on two NVIDIA GeForce RTX 3090 GPU machines. DynEformer contains 2 GP-encoder blocks and 1 decoder layer.



\textbf{Evaluation metric \& baselines.} 
We use two evaluation metrics, including MSE $=\frac{1}{n} \sum_{i=1}^n(\mathcal{Y}-\hat{\mathcal{Y}})^2$ and MAE $=\frac{1}{n} \sum_{i=1}^n|\mathcal{Y}-\hat{\mathcal{Y}}|$. 

We select six time-series prediction methods as comparison, including three latest state-of-the-art transformer-based models: Deep Transformer\cite{wu2020deep}, Informer\cite{Zhou2021}, and Autoformer\cite{wu2021autoformer}; two RNN-based models: MQRNN \cite{Wen2017} and DeepAR \cite{SALINAS20201181}; and a clustering-based model: VaRDE-LSTM \cite{9488816}. These models are generalized to solve most classical time-series forecasting problems, and some perform well on similar power and traffic workload forecasting tasks. For fairness, static context data are input as marks when baselines are implemented on ECW datasets.

Considering the model training costs, we identify five data classes and trained five separate predictors for VaRDE-LSTM (VaRDE-L) on the two datasets (resulting in 3.5× training time compared to the other models).



\subsection{Global Pooling}\label{GPRes}
\begin{figure}[htbp]
    \centering
    \setlength{\belowcaptionskip}{-0.3cm}
    \subfigure[ECW]{\label{fig:ecw_gp}
    \includegraphics[width=4.0cm]{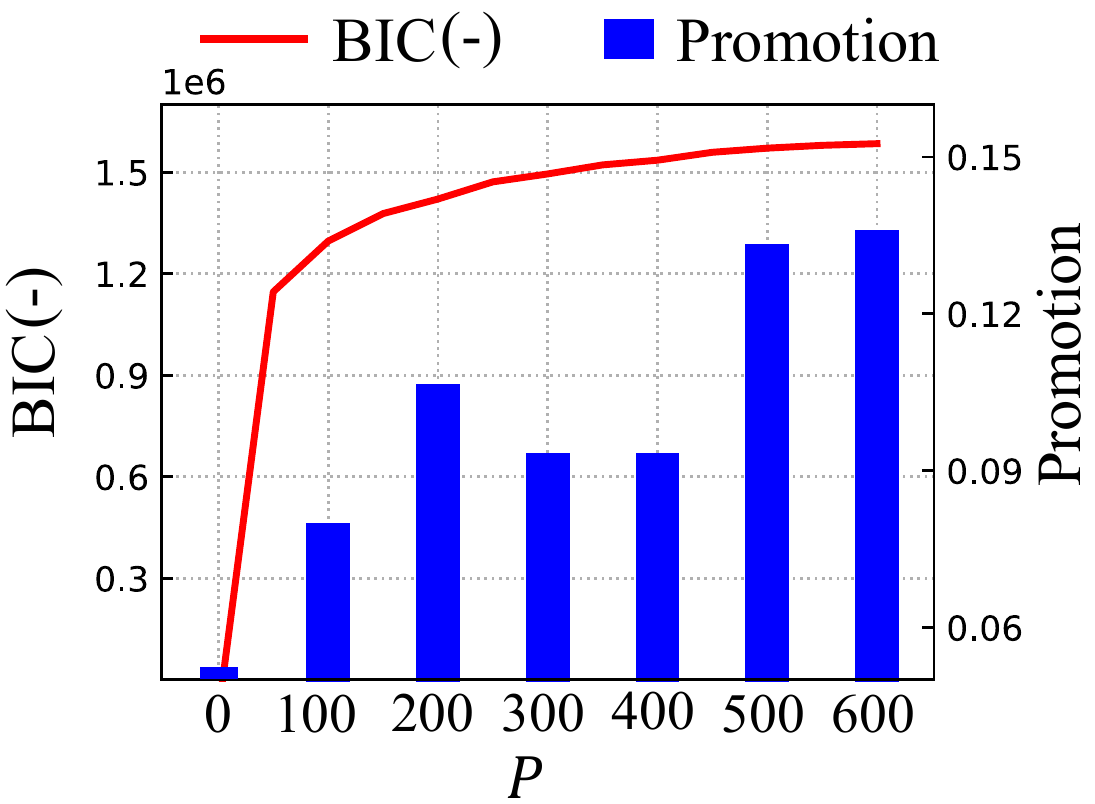}
    }
    \subfigure[Azure]{\label{fig:azure_gp}
    \includegraphics[width=4.0cm]{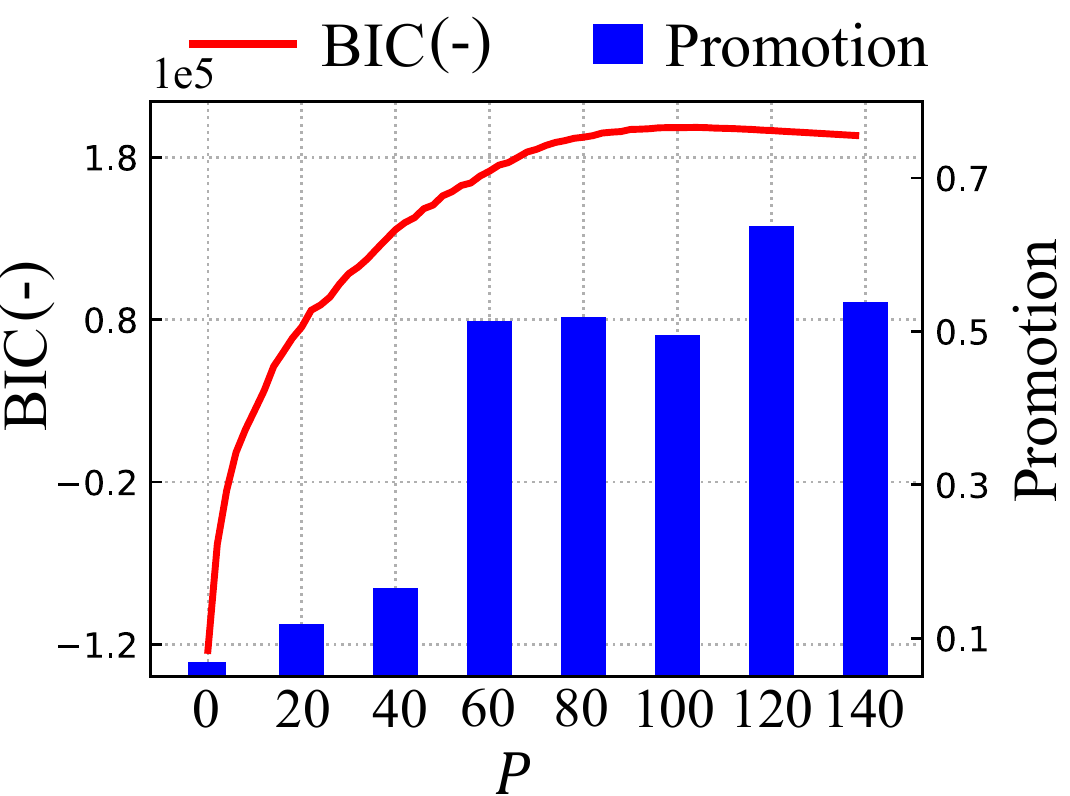}
    }
    \caption{The BIC and performance promotion at different $P$, where BIC(-) represents the negative of BIC, the larger the better, and promotion is obtained by comparing the MSE decrease of the no-global pooling model.}
    \label{fig:gp promotion}
\end{figure}

To determine the optimal global pool size $P$, we obtain the BIC of the VaDE model for different $P$ and give the percentage of performance promotion of the DynEformer for different sizes of global pools based on the validation set. The results are reported in Fig. \ref{fig:gp promotion}.
\vspace{-0.3cm}

As Fig. \ref{fig:gp promotion} shows, the BIC of VaDE's clustering results decreases as $P$ increases, while the performance promotion of the global pool on DynEformer follows the increasing trend. On ECW, the BIC decreases significantly as $P$ increases from 0 to 400 and converges after $P=400$ while the performance promotion has two abrupt changes at $P=200$ and $P=500$, indicating that the global pool extracts the global application patterns sufficiently in these settings. Therefore, we set the global pool of $P=500$ for the DynEformer on ECW. Likewise, the global pool on Azure is set to $P=120$.

To conclude, choosing $P$ for maximum BIC(-) led to similar performance as MSE promotion. In real-world scenarios, we rely on BIC (obtained during VaDE clustering) to select the optimal $P$, avoiding exposure of a priori information from the future.


\begin{figure}[htbp]
    \centering
    \setlength{\belowcaptionskip}{-0.3cm}
    \includegraphics[width=\columnwidth]{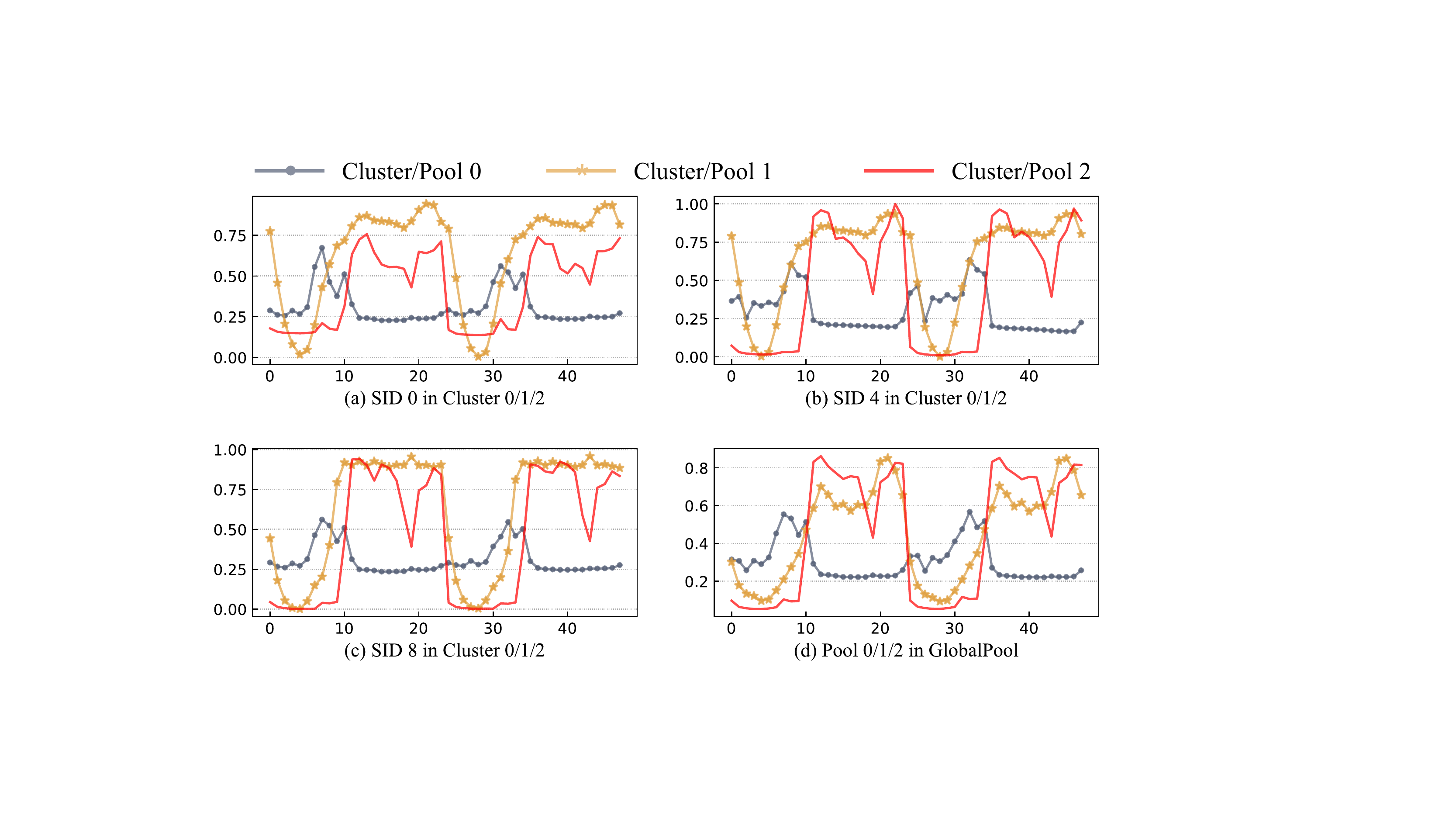}
    \caption{Clusters and global pool samples of ECW.}
    \label{fig:Global Pool}
\end{figure}

Fig. \ref{fig:Global Pool} shows the results of VaDE's clustering and the global pool for the workloads' seasonal component in the ECW. Where Fig. \ref{fig:Global Pool}(a), (b), and (c) are different workload series samples in clusters 0, 1, 2, respectively, and Fig. \ref{fig:Global Pool} (d) is the corresponding  0th, 1st and 2nd pool in the global pool.

As shown in Fig. \ref{fig:Global Pool}(a)-(c), series in the same clusters have similar patterns, which are consistent with the pattern of the corresponding pool in Fig. \ref{fig:Global Pool}(d). The results indicate that VaDE effectively aggregate the global workload patterns of applications, thus capturing the workload similarity of the same application type.



\subsection{Edge Clouds Workload Prediction}\label{ExpR}


\textbf{Overall-evaluation.} From Table \ref{table:main result}, we can observe that: (\textbf{1}) DynEformer achieves the consistent state-of-the-art performance in all benchmarks across the two workload datasets. Especially, compared to the transformer-based models family, DynEformer decreases MSE by an average of 12\% in ECW and 83\% in Azure. DynEformer outperforms MQRNN and DeepAR on MSE by decreasing 18\%, 95\% (in ECW), 97\%, and 99\% (in Azure). Compared to the clustering-based model, DynEformer gives 55\% (0.150→0.067, in ECW) and 94\% (1.124→0.069, in Azure) MSE reduction.  

(\textbf{2}) The performance of transformer family is not guaranteed to outperform RNN-based model. In fact, as seen in Table \ref{table:main result}, MQRNN outperforms Autoformer and Informer in Azure. This reveals in short-term workload prediction scenarios, the various optimizations performed for the self-attention mechanism may not lead to better results than RNN models with long network paths. However, DynEformer still outperforms MQRNN and Deep Transformer, which reflects the unsurpassed prediction capacity of DynEformer.




\textbf{Robustness to app switching (concept drift).}  As Table \ref{table:main result} shows, the performance of all the baseline models decreased dramatically on application switching data picked out from ECW (100\%, 36\%, 95\%, 309\% and  80\% MSE raised for Autoformer, Informer, Deep Trans., MQRNN and VaRDE-L separately). Among them, our method shows the lowest error, and impressively, DynEformer maintains the consistent performance on the MSE metric, and the MAE is 20\% raised. The result shows that the proposed method can robustly predict even under MT-ECP application switching.

\begin{figure*}[htbp]
\centering
\subfigure[Work Steadily]{\label{fig:normal}
\includegraphics[height=3.2cm, width=6.47cm]{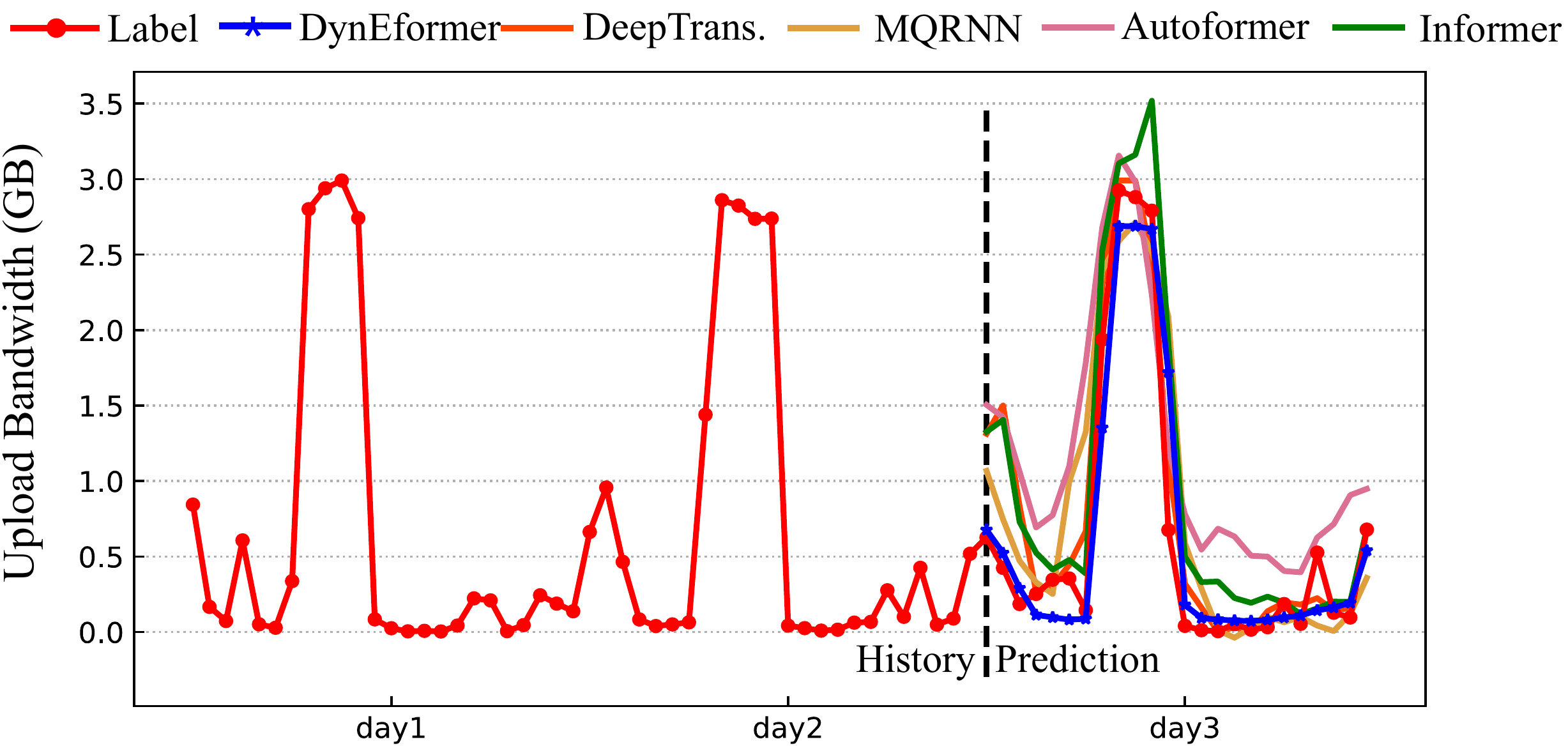}
}
\hspace{0.5cm}
\subfigure[App Switch]{\label{fig:switch}
\includegraphics[height=3.2cm, width=6.47cm]{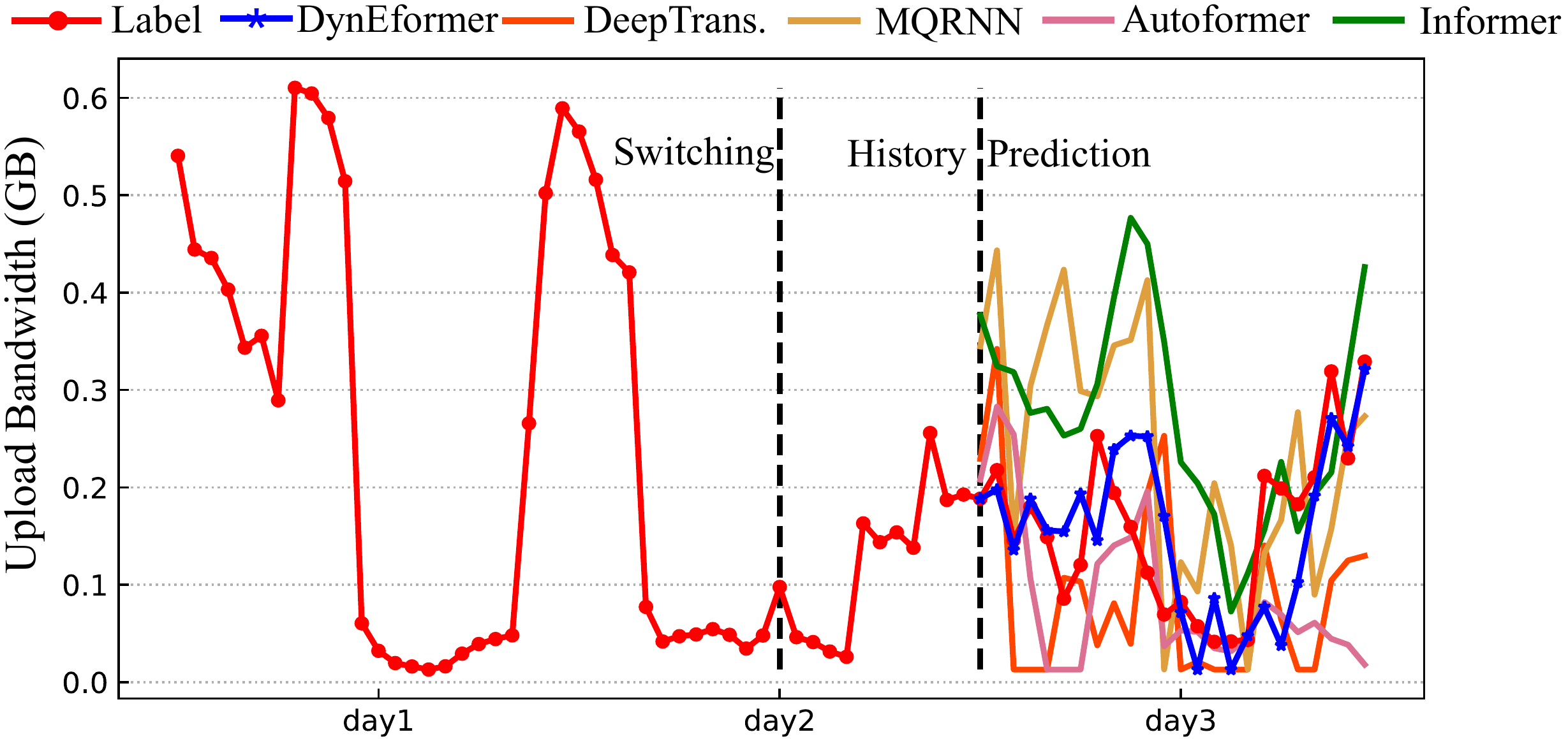}
}
\subfigure[New Infrastructure]{\label{fig:newmac}
\includegraphics[height=3.2cm, width=6.47cm]{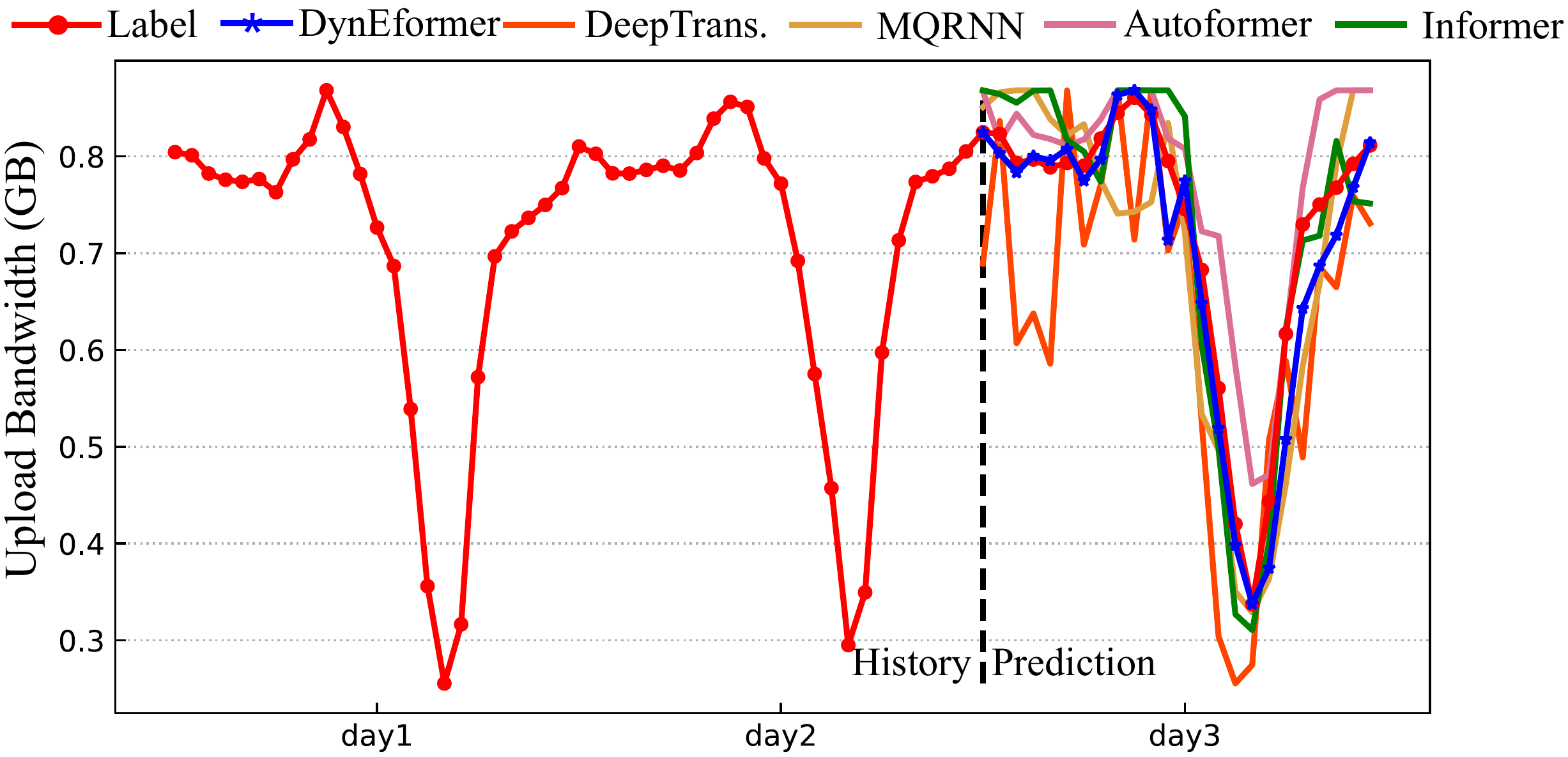}
}
\hspace{0.5cm}
\subfigure[New App]{\label{fig:newapp}
\includegraphics[height=3.2cm, width=6.47cm]{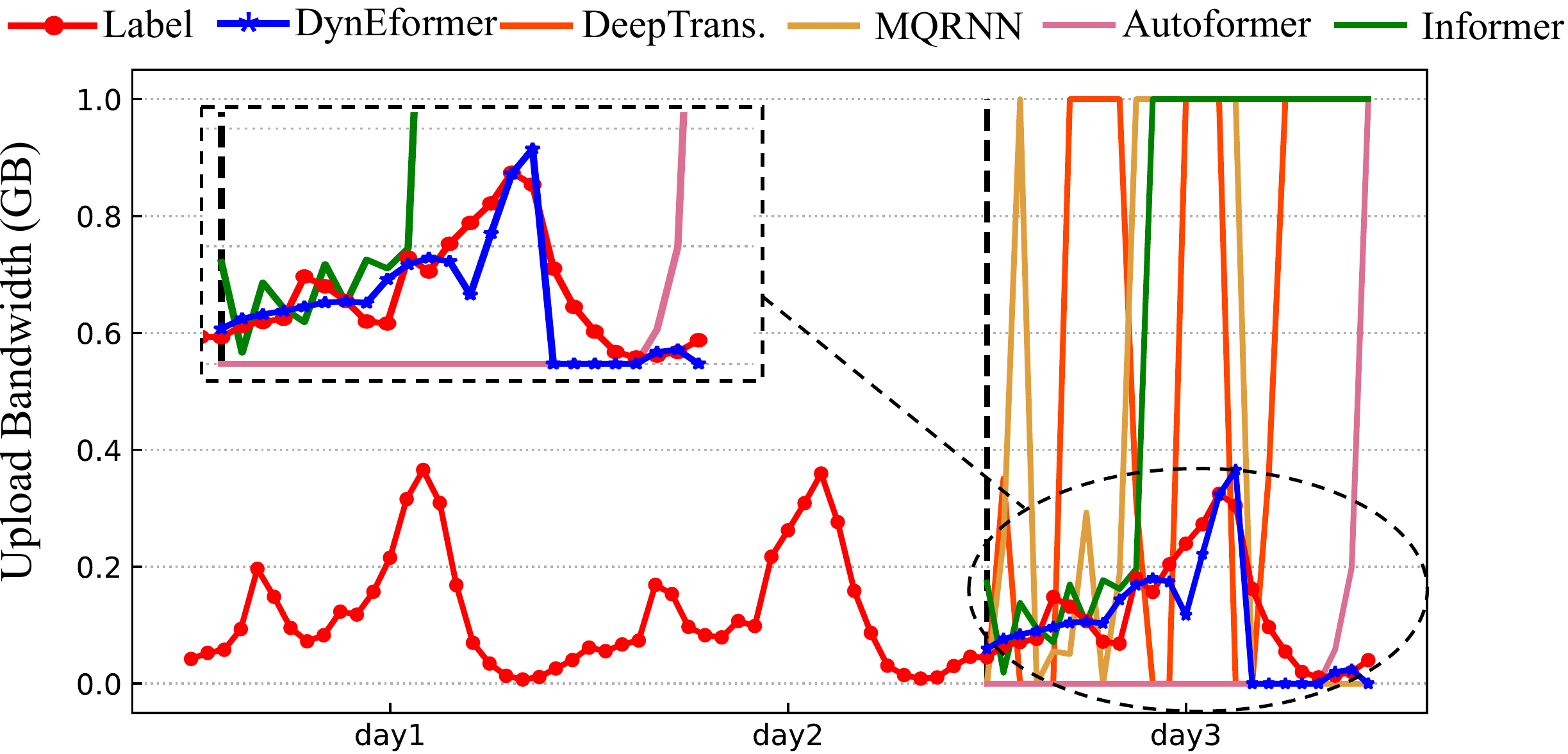}
}
\caption{Prediction results for different MT-ECP behaviors.}
\end{figure*}

As can be seen in Fig. \ref{fig:switch}, the application switching that occurs on day two results in an abrupt change in the workload distribution, i.e., an abrupt-style concept drift. However, even with only 1/4 of the valid post-switch inputs, DynEformer still captures the correct workload pattern and diminishes the impact of the before-switch workload pattern and magnitude. This is aided by the DynEformer's effective capture of global pattern information, which allows it to accurately detect subtle differences between the local workload and the regular global pattern. As these processes occur covertly within the model weights, users do not need to be concerned about the additional cost of updating and switching the model.


\textbf{Robustness to unseen entities (cold start).} The DynEformer is robust to unseen entities, leading to high scalability. We evaluate DynEformer for unseen application and infrastructure data via the ECW-New Infras. and ECW-New App. As can be seen from Table \ref{table:main result}, DynEformer outperforms all baselines for the two unseen entities-oriented data. On average, DynEformer outperforms the transformers family on ECW-New Infras. by decreasing 10\% MSE and outperforms MQRNN and VaRDE-L by decreasing 15\% and 72\% MSE. In addition, on the more challenging ECW-New App, the MSE decrease is further improved to 32\%, 94\%, and 84\%.


As can be seen in Fig. \ref{fig:newmac} and Fig. \ref{fig:newapp}, even one may not feel any disharmony with the prediction result via the DynEformer model compared to other approaches. This result shows that the proposed framework can perform optimal prediction on the data distribution that has not been seen in training without additional learning. Therefore, the framework is scalable when new infrastructure and applications are added.


\vspace{-0.3cm}
\begin{figure}[htbp]
    \centering
    \includegraphics[width=8cm]{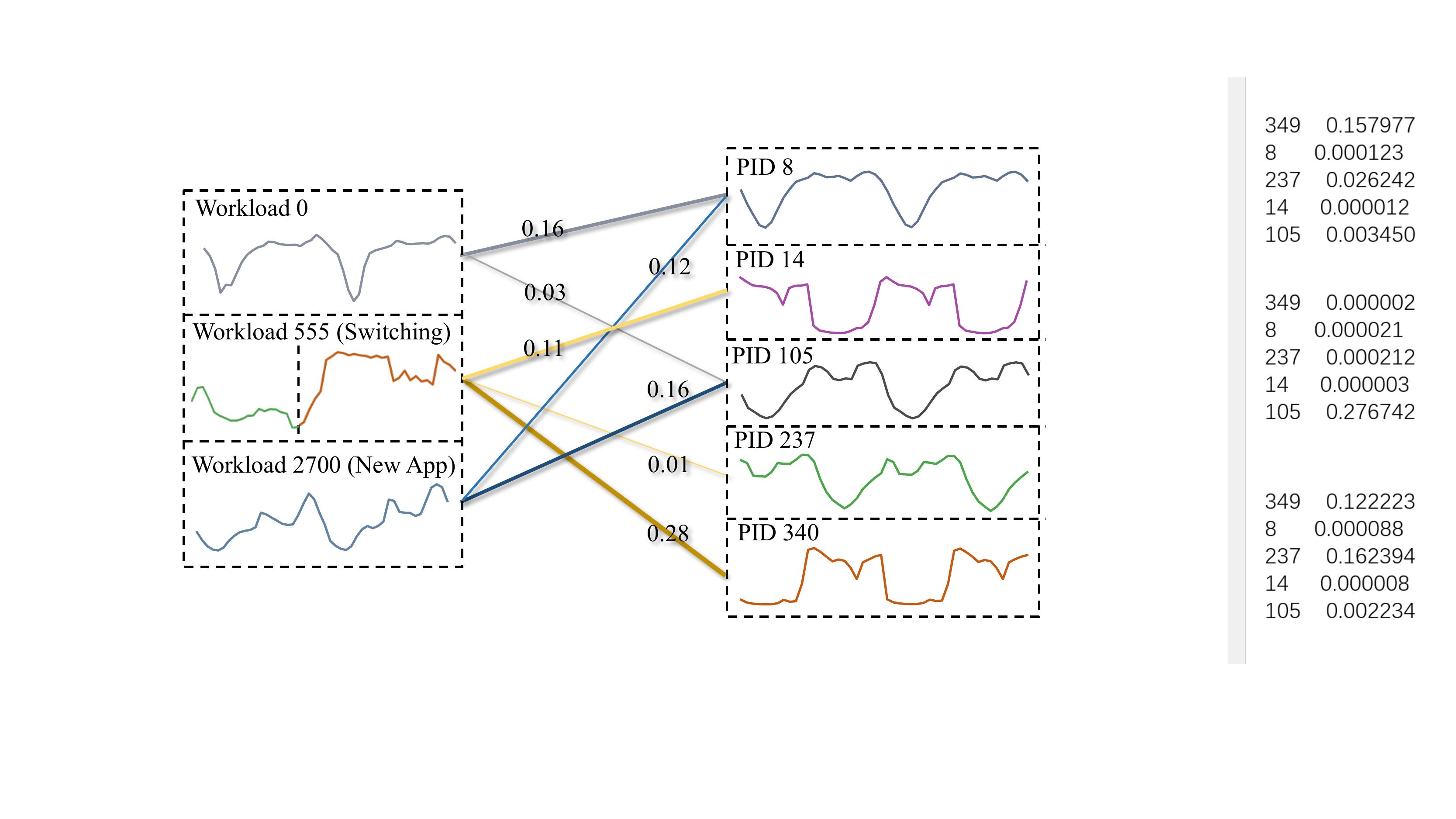}
    \caption{An example of the global-pool-merge mechanism in the GP-encoder. Instead of creating separate models for each cluster, the GP layer can choose the most suitable global pool for each local workload. Best viewed in color.}
    \label{fig:gp visualization}
\end{figure}

\begin{table}[t]
\setlength{\belowcaptionskip}{-0.3cm}
\caption{Ablation of DynEformer in all ECW datasets with MAE metric. DynEformer(-S). removes the static content and the SA layer from the DynEformer, DynEformer(-GP) removes the global pooling and the GP layer, 0Padding replaces the synchronous padding with the zero constant. 'Promotion' and '(x\%)' represents the MAE improvement of the DynEformer on the corresponding dataset or model.}
\label{table:ablation}
\resizebox{\columnwidth}{!}{
\begin{tabular}{@{}l|cccc@{}}
\toprule
Models      & DynEformer & DynEformer(-GP) & DynEformer(-S) & 0Padding     \\ \midrule
ECW         & 0.141     & 0.157(10\%)  & 0.149(5\%)   & 0.152(7\%)   \\ \midrule
App Switch  & 0.168     & 0.199(16\%)  & 0.185(9\%)   & 0.185(9\%)   \\ \midrule
New Infras. & 0.143     & 0.155(7\%)   & 0.151(5\%)   & 0.158(9\%) \\ \midrule
New App     & 0.207     & 0.230(13\%)  & 0.207(0\%)   & 0.214(3\%)   \\ \midrule
Promotion   & -         & 12\%         & 5\%          & 8\%          \\ \bottomrule
\end{tabular}}
\end{table}

\subsection{How DynEformer works: Ablation Study and Model Analysis}
In this section, we perform ablation studies to illustrate the role of different components. As can be seen from Table \ref{table:ablation}, each component and mechanism in DynEformer has improved the model's prediction performance under dynamic MT-ECP behaviors. The global pooling gives a combined improvement of 12\% over the DynEformer(-GP), which indicates that the global pooling effectively solves the problems of complex patterns, concept drift, and cold start caused by dynamic MT-ECP behaviors and improves the applicability of the transformer in real-world predictions.


The SA layer plays a significant role in the MT-ECP behaviors of application switching and new infrastructure addition, bringing 9\% and 5\% to the model improvement, respectively, which indicates that DynEformer effectively exploits the static infrastructure attributes to combat dynamic instability in workload prediction.


Furthermore, the 0Padding model performs worse than the DynEformer in all situations, although it remains the GP layer and the SA layer. The effect of the global pooling is diminished when the synchronous padding is deactivated because the added global information does not match the 0 padding and misleads the attached encoder-decoder attention.


Fig. \ref{fig:gp visualization} shows how well GP layer works. With GP layer, local workloads can be associated with the most relevant global application patterns. For example, the workload with application switching (workload 555) is associated with the global pool with the before-switch application pattern (PID 237) and the post-switch application pattern (PID 14 and 340) by the GP layer. Further, the GP layer successfully focuses its attention on the global pattern with PID 340 at weight 0.28 and less on PID 237 at weight 0.01. Therefore, the DynEformer can prevent being misled by lagging local patterns and preemptively switch its attention to the new pattern.

In another instance, the GP layer successfully identifies the new workload pattern (workload 2700) by overlaying two existing global pools (PID 8 and PID 105). The results show that our proposed model can effectively merge global workload patterns for local workload prediction and alleviate the distraction caused by intricate patterns when application switching and new entities are added.


\begin{table}[htbp]
\caption{Comparison of application depreciation rate rank. '\#' indicates the depreciation rate rank, the \textbf{bold} indicates wrong ranks and 'Count' represents the correct ranks number.}
\label{table:use_case}
\centering
\begin{tabular}{l|ccc}
\hline
Application        & Label & DynEformer & Informer           \\ \midrule
$APP_1$        & 0.075  \#1 & 0.049 \#1 & \textbf{0.024 \#3} \\ \midrule
$APP_2$        & 0.068  \#2 & 0.044 \#2 & \textbf{0.059 \#1} \\ \midrule
$APP_3$        & 0.033  \#3 & 0.017 \#3 & \textbf{0.041 \#2} \\ \midrule
$APP_4$        & 0.011  \#4 & 0.015 \#4 & 0.013 \#4          \\ \midrule
$APP_5$        & 0.002  \#5 & 0.001 \#5 & 0.001 \#5         \\ \bottomrule
Count          & -          & 5         & 2         \\ \hline
\end{tabular}
\end{table}

 \subsection{Use Case in MT-ECP}  

We compare the best-performing baseline (Informer) and DynEform-er for the test period (08/25-08/30) of ECW in terms of application depreciation rate, which can be obtained from the workload series, and rank them from largest to smallest. 


The application depreciation rate is an important metric in MT-ECP to measure application revenue. Application depreciation comes from the difference in billing time between the application provider and the device provider, i.e., the billing application workload is less than the sum of the billing workloads of all devices running the application, resulting in less revenue workload than expense workload on that application. The application $a$'s depreciation rate $D_a$ can be formalized as follows:
\begin{equation}
    D_a = 1-\frac{\sum_{l=1}^{k}x_{t_a}^l}{\sum_{l=1}^{k}x_{t_l}^l}, \mathbf{App}(x^l) = a
\end{equation}where $x_{t_a}^l$ is the workload at the application billing time $t_a$, $x_{t_l}^l$ is the workload at different device billing times $t_l$.

The results are reported in table \ref{table:use_case}. The DynEformer beats Informer mostly in correct rank counts, i.e., 5>2, which supports DynEformer's potential of assisting in decision making in MT-ECP. Looking at the result of the Informer, one may be led to believe that the depreciation rate of $APP_2$ and $APP_3$ is larger than the $APP_1$. Since the higher depreciation rate means lower revenue efficiency, this will lead the MT-ECP to reduce the deployment of these applications. However, the actual ranking tells a very different story—one where the depreciation rate of $APP_2$ and $APP_3$ is less than that of $APP_1$, and more infrastructure should be deployed with $APP_3$ compared to $APP_1$\footnote{In fact, CPs have certain agreements with MT-ECP regarding the range of workload demand, which also provides a scheduling space for MT-ECP.}. DynEformer successfully arrives at the correct conclusion and thus avoids a huge loss of revenue.



\section{Conclusion}\label{conclusion}
We proposed DynEformer, a transformer for workload prediction in dynamic MT-ECP employing global pooling and static context awareness. The global pool, built on identified workload factors, improves local workload prediction through a new information merging mechanism. Experimental results show DynEformer surpasses current standards on five practical datasets. The ablation study highlights the essence and effectiveness of our design choices. The use case in real MT-ECP shows how DynEformer can support decisions and improve the benefits of application deployment.

We believe that DynEformer is an effective alternative to clustering based approaches, incorporating former's capability to capture kindred patterns and extending its application to a more flexible way. In the future we will explore the auto-update mechanisms for global pooling so that users can use DynEformer for long-term prediction without intermittently updating the global pool.

\begin{acks}
Thanks to Paiou Cloud Computing (Shanghai) Co., Ltd for providing the system platform and original logs. This work was supported in part by China NSFC through grant No. 62072332 and China NSFC (Youth) through grant No. 62002260; in part by the Tianjin Xinchuang Haihe Lab under Grant No.22HHXCJC00002.
\end{acks}

\bibliographystyle{ACM-Reference-Format}
\balance
\bibliography{sample-base}


\begin{thebibliography}{31}


\ifx \showCODEN    \undefined \def \showCODEN     #1{\unskip}     \fi
\ifx \showDOI      \undefined \def \showDOI       #1{#1}\fi
\ifx \showISBNx    \undefined \def \showISBNx     #1{\unskip}     \fi
\ifx \showISBNxiii \undefined \def \showISBNxiii  #1{\unskip}     \fi
\ifx \showISSN     \undefined \def \showISSN      #1{\unskip}     \fi
\ifx \showLCCN     \undefined \def \showLCCN      #1{\unskip}     \fi
\ifx \shownote     \undefined \def \shownote      #1{#1}          \fi
\ifx \showarticletitle \undefined \def \showarticletitle #1{#1}   \fi
\ifx \showURL      \undefined \def \showURL       {\relax}        \fi
\providecommand\bibfield[2]{#2}
\providecommand\bibinfo[2]{#2}
\providecommand\natexlab[1]{#1}
\providecommand\showeprint[2][]{arXiv:#2}

\bibitem[Arbat et~al\mbox{.}(2022)]%
        {arbat2022wasserstein}
\bibfield{author}{\bibinfo{person}{Shivani Arbat},
  \bibinfo{person}{Vinodh~Kumaran Jayakumar}, \bibinfo{person}{Jaewoo Lee},
  \bibinfo{person}{Wei Wang}, {and} \bibinfo{person}{In~Kee Kim}.}
  \bibinfo{year}{2022}\natexlab{}.
\newblock \showarticletitle{Wasserstein Adversarial Transformer for Cloud
  Workload Prediction}. In \bibinfo{booktitle}{\emph{Thirty-Sixth {AAAI}
  Conference on Artificial Intelligence, {AAAI} 2022}}.
  \bibinfo{publisher}{{AAAI} Press}, \bibinfo{pages}{12433--12439}.
\newblock


\bibitem[Azmandian et~al\mbox{.}(2011)]%
        {6005369}
\bibfield{author}{\bibinfo{person}{Fatemeh Azmandian}, \bibinfo{person}{Micha
  Moffie}, \bibinfo{person}{Jennifer~G. Dy}, \bibinfo{person}{Javed~A. Aslam},
  {and} \bibinfo{person}{David~R. Kaeli}.} \bibinfo{year}{2011}\natexlab{}.
\newblock \showarticletitle{Workload Characterization at the Virtualization
  Layer}. In \bibinfo{booktitle}{\emph{2011 IEEE 19th Annual International
  Symposium on Modelling, Analysis, and Simulation of Computer and
  Telecommunication Systems}}. \bibinfo{pages}{63--72}.
\newblock


\bibitem[Calheiros et~al\mbox{.}(2015)]%
        {6881647}
\bibfield{author}{\bibinfo{person}{Rodrigo~N. Calheiros},
  \bibinfo{person}{Enayat Masoumi}, \bibinfo{person}{Rajiv Ranjan}, {and}
  \bibinfo{person}{Rajkumar Buyya}.} \bibinfo{year}{2015}\natexlab{}.
\newblock \showarticletitle{Workload Prediction Using ARIMA Model and Its
  Impact on Cloud Applications’ QoS}.
\newblock \bibinfo{journal}{\emph{IEEE Transactions on Cloud Computing}}
  \bibinfo{volume}{3}, \bibinfo{number}{4} (\bibinfo{year}{2015}),
  \bibinfo{pages}{449--458}.
\newblock


\bibitem[Cortez et~al\mbox{.}(2017)]%
        {10.1145/3132747.3132772}
\bibfield{author}{\bibinfo{person}{Eli Cortez}, \bibinfo{person}{Anand Bonde},
  \bibinfo{person}{Alexandre Muzio}, \bibinfo{person}{Mark Russinovich},
  \bibinfo{person}{Marcus Fontoura}, {and} \bibinfo{person}{Ricardo
  Bianchini}.} \bibinfo{year}{2017}\natexlab{}.
\newblock \showarticletitle{Resource Central: Understanding and Predicting
  Workloads for Improved Resource Management in Large Cloud Platforms}. In
  \bibinfo{booktitle}{\emph{Proceedings of the 26th Symposium on Operating
  Systems Principles}} (Shanghai, China) \emph{(\bibinfo{series}{SOSP '17})}.
  \bibinfo{publisher}{Association for Computing Machinery},
  \bibinfo{address}{New York, NY, USA}, \bibinfo{pages}{153–167}.
\newblock
\showISBNx{9781450350853}


\bibitem[Cui et~al\mbox{.}(2022)]%
        {10.14778/3489496.3489503}
\bibfield{author}{\bibinfo{person}{Yue Cui}, \bibinfo{person}{Kai Zheng},
  \bibinfo{person}{Dingshan Cui}, \bibinfo{person}{Jiandong Xie},
  \bibinfo{person}{Liwei Deng}, \bibinfo{person}{Feiteng Huang}, {and}
  \bibinfo{person}{Xiaofang Zhou}.} \bibinfo{year}{2022}\natexlab{}.
\newblock \showarticletitle{METRO: A Generic Graph Neural Network Framework for
  Multivariate Time Series Forecasting}.
\newblock \bibinfo{journal}{\emph{Proc. VLDB Endow.}} \bibinfo{volume}{15},
  \bibinfo{number}{2} (\bibinfo{date}{feb} \bibinfo{year}{2022}),
  \bibinfo{pages}{224–236}.
\newblock
\showISSN{2150-8097}


\bibitem[Duc et~al\mbox{.}(2019)]%
        {10.1145M}
\bibfield{author}{\bibinfo{person}{Thang~Le Duc},
  \bibinfo{person}{Rafael~Garc\'{\i}a Leiva}, \bibinfo{person}{Paolo Casari},
  {and} \bibinfo{person}{Per-Olov \"{O}stberg}.}
  \bibinfo{year}{2019}\natexlab{}.
\newblock \showarticletitle{Machine Learning Methods for Reliable Resource
  Provisioning in Edge-Cloud Computing: A Survey}.
\newblock \bibinfo{journal}{\emph{ACM Comput. Surv.}}, Article
  \bibinfo{articleno}{94} (\bibinfo{date}{sep} \bibinfo{year}{2019}),
  \bibinfo{numpages}{39}~pages.
\newblock


\bibitem[Gao et~al\mbox{.}(2020)]%
        {9209730}
\bibfield{author}{\bibinfo{person}{Jiechao Gao}, \bibinfo{person}{Haoyu Wang},
  {and} \bibinfo{person}{Haiying Shen}.} \bibinfo{year}{2020}\natexlab{}.
\newblock \showarticletitle{Machine Learning Based Workload Prediction in Cloud
  Computing}. In \bibinfo{booktitle}{\emph{2020 29th International Conference
  on Computer Communications and Networks (ICCCN)}}. \bibinfo{pages}{1--9}.
\newblock


\bibitem[Jayakumar et~al\mbox{.}(2020)]%
        {jayakumar2020self}
\bibfield{author}{\bibinfo{person}{Vinodh~Kumaran Jayakumar},
  \bibinfo{person}{Jaewoo Lee}, \bibinfo{person}{In~Kee Kim}, {and}
  \bibinfo{person}{Wei Wang}.} \bibinfo{year}{2020}\natexlab{}.
\newblock \showarticletitle{A Self-Optimized Generic Workload Prediction
  Framework for Cloud Computing}. In \bibinfo{booktitle}{\emph{2020 IEEE
  International Parallel and Distributed Processing Symposium (IPDPS)}}. IEEE,
  \bibinfo{pages}{779--788}.
\newblock


\bibitem[Jia et~al\mbox{.}(2014)]%
        {6983058}
\bibfield{author}{\bibinfo{person}{Zhen Jia}, \bibinfo{person}{Jianfeng Zhan},
  \bibinfo{person}{Lei Wang}, \bibinfo{person}{Rui Han},
  \bibinfo{person}{Sally~A. McKee}, \bibinfo{person}{Qiang Yang},
  \bibinfo{person}{Chunjie Luo}, {and} \bibinfo{person}{Jingwei Li}.}
  \bibinfo{year}{2014}\natexlab{}.
\newblock \showarticletitle{Characterizing and subsetting big data workloads}.
  In \bibinfo{booktitle}{\emph{2014 IEEE International Symposium on Workload
  Characterization (IISWC)}}. \bibinfo{pages}{191--201}.
\newblock


\bibitem[Jiang et~al\mbox{.}(2017)]%
        {ijcai2017p273}
\bibfield{author}{\bibinfo{person}{Zhuxi Jiang}, \bibinfo{person}{Yin Zheng},
  \bibinfo{person}{Huachun Tan}, \bibinfo{person}{Bangsheng Tang}, {and}
  \bibinfo{person}{Hanning Zhou}.} \bibinfo{year}{2017}\natexlab{}.
\newblock \showarticletitle{Variational Deep Embedding: An Unsupervised and
  Generative Approach to Clustering}. In \bibinfo{booktitle}{\emph{Proceedings
  of the Twenty-Sixth International Joint Conference on Artificial
  Intelligence, {IJCAI-17}}}. \bibinfo{pages}{1965--1972}.
\newblock


\bibitem[Kim et~al\mbox{.}(2018)]%
        {8457781}
\bibfield{author}{\bibinfo{person}{In~Kee Kim}, \bibinfo{person}{Wei Wang},
  \bibinfo{person}{Yanjun Qi}, {and} \bibinfo{person}{Marty Humphrey}.}
  \bibinfo{year}{2018}\natexlab{}.
\newblock \showarticletitle{CloudInsight: Utilizing a Council of Experts to
  Predict Future Cloud Application Workloads}. In
  \bibinfo{booktitle}{\emph{2018 IEEE 11th International Conference on Cloud
  Computing (CLOUD)}}. \bibinfo{pages}{41--48}.
\newblock


\bibitem[Kingma and Ba(2014)]%
        {kingma2014adam}
\bibfield{author}{\bibinfo{person}{Diederik~P Kingma} {and}
  \bibinfo{person}{Jimmy Ba}.} \bibinfo{year}{2014}\natexlab{}.
\newblock \showarticletitle{Adam: A method for stochastic optimization}.
\newblock \bibinfo{journal}{\emph{arXiv preprint arXiv:1412.6980}}
  (\bibinfo{year}{2014}).
\newblock


\bibitem[Le~Duc and Oestberg(2018)]%
        {8487450}
\bibfield{author}{\bibinfo{person}{Thang Le~Duc} {and}
  \bibinfo{person}{Per-Olov Oestberg}.} \bibinfo{year}{2018}\natexlab{}.
\newblock \showarticletitle{Application, Workload, and Infrastructure Models
  for Virtualized Content Delivery Networks Deployed in Edge Computing
  Environments}. In \bibinfo{booktitle}{\emph{2018 27th International
  Conference on Computer Communication and Networks (ICCCN)}}.
  \bibinfo{pages}{1--7}.
\newblock


\bibitem[Lim et~al\mbox{.}(2021)]%
        {LIM20211748}
\bibfield{author}{\bibinfo{person}{Bryan Lim}, \bibinfo{person}{Sercan~Ö.
  Arık}, \bibinfo{person}{Nicolas Loeff}, {and} \bibinfo{person}{Tomas
  Pfister}.} \bibinfo{year}{2021}\natexlab{}.
\newblock \showarticletitle{Temporal Fusion Transformers for interpretable
  multi-horizon time series forecasting}.
\newblock \bibinfo{journal}{\emph{International Journal of Forecasting}}
  \bibinfo{volume}{37}, \bibinfo{number}{4} (\bibinfo{year}{2021}),
  \bibinfo{pages}{1748--1764}.
\newblock
\showISSN{0169-2070}


\bibitem[Liu et~al\mbox{.}(2016)]%
        {7562213}
\bibfield{author}{\bibinfo{person}{Bingwei Liu}, \bibinfo{person}{Yinan Lin},
  {and} \bibinfo{person}{Yu Chen}.} \bibinfo{year}{2016}\natexlab{}.
\newblock \showarticletitle{Quantitative workload analysis and prediction using
  Google cluster traces}. In \bibinfo{booktitle}{\emph{2016 IEEE Conference on
  Computer Communications Workshops (INFOCOM WKSHPS)}}.
  \bibinfo{pages}{935--940}.
\newblock


\bibitem[Liu et~al\mbox{.}(2017)]%
        {LIU201735}
\bibfield{author}{\bibinfo{person}{Chunhong Liu}, \bibinfo{person}{Chuanchang
  Liu}, \bibinfo{person}{Yanlei Shang}, \bibinfo{person}{Shiping Chen},
  \bibinfo{person}{Bo Cheng}, {and} \bibinfo{person}{Junliang Chen}.}
  \bibinfo{year}{2017}\natexlab{}.
\newblock \showarticletitle{An adaptive prediction approach based on workload
  pattern discrimination in the cloud}.
\newblock \bibinfo{journal}{\emph{Journal of Network and Computer
  Applications}}  \bibinfo{volume}{80} (\bibinfo{year}{2017}),
  \bibinfo{pages}{35--44}.
\newblock
\showISSN{1084-8045}


\bibitem[Lu et~al\mbox{.}(2019)]%
        {8496795}
\bibfield{author}{\bibinfo{person}{Jie Lu}, \bibinfo{person}{Anjin Liu},
  \bibinfo{person}{Fan Dong}, \bibinfo{person}{Feng Gu}, \bibinfo{person}{João
  Gama}, {and} \bibinfo{person}{Guangquan Zhang}.}
  \bibinfo{year}{2019}\natexlab{}.
\newblock \showarticletitle{Learning under Concept Drift: A Review}.
\newblock \bibinfo{journal}{\emph{IEEE Transactions on Knowledge and Data
  Engineering}} \bibinfo{volume}{31}, \bibinfo{number}{12}
  (\bibinfo{year}{2019}), \bibinfo{pages}{2346--2363}.
\newblock


\bibitem[Mumolo et~al\mbox{.}(2017)]%
        {Mumolo2017ErgodicHM}
\bibfield{author}{\bibinfo{person}{Enzo Mumolo}, \bibinfo{person}{Gianni~Viardo
  Vercelli}, {and} \bibinfo{person}{Alfredo Cuzzocrea}.}
  \bibinfo{year}{2017}\natexlab{}.
\newblock \showarticletitle{Ergodic Hidden Markov Models for Workload
  Characterization Problems}. In \bibinfo{booktitle}{\emph{DMSVLSS}}.
\newblock


\bibitem[Paszke et~al\mbox{.}(2019)]%
        {paszke2019pytorch}
\bibfield{author}{\bibinfo{person}{Adam Paszke}, \bibinfo{person}{Sam Gross},
  \bibinfo{person}{Francisco Massa}, \bibinfo{person}{Adam Lerer},
  \bibinfo{person}{James Bradbury}, \bibinfo{person}{Gregory Chanan},
  \bibinfo{person}{Trevor Killeen}, \bibinfo{person}{Zeming Lin},
  \bibinfo{person}{Natalia Gimelshein}, \bibinfo{person}{Luca Antiga},
  {et~al\mbox{.}}} \bibinfo{year}{2019}\natexlab{}.
\newblock \showarticletitle{Pytorch: An imperative style, high-performance deep
  learning library}.
\newblock \bibinfo{journal}{\emph{Advances in neural information processing
  systems}}  \bibinfo{volume}{32} (\bibinfo{year}{2019}).
\newblock


\bibitem[Salinas et~al\mbox{.}(2020)]%
        {SALINAS20201181}
\bibfield{author}{\bibinfo{person}{David Salinas}, \bibinfo{person}{Valentin
  Flunkert}, \bibinfo{person}{Jan Gasthaus}, {and} \bibinfo{person}{Tim
  Januschowski}.} \bibinfo{year}{2020}\natexlab{}.
\newblock \showarticletitle{DeepAR: Probabilistic forecasting with
  autoregressive recurrent networks}.
\newblock \bibinfo{journal}{\emph{International Journal of Forecasting}}
  \bibinfo{volume}{36}, \bibinfo{number}{3} (\bibinfo{year}{2020}),
  \bibinfo{pages}{1181--1191}.
\newblock
\showISSN{0169-2070}


\bibitem[Sun et~al\mbox{.}(2021)]%
        {SUN2021107625}
\bibfield{author}{\bibinfo{person}{Linjin Sun}, \bibinfo{person}{Yangjian Ji},
  \bibinfo{person}{Mingrui Zhu}, \bibinfo{person}{Fu Gu}, \bibinfo{person}{Feng
  Dai}, {and} \bibinfo{person}{Ke Li}.} \bibinfo{year}{2021}\natexlab{}.
\newblock \showarticletitle{A new predictive method supporting streaming data
  with hybrid recurring concept drifts in process industry}.
\newblock \bibinfo{journal}{\emph{Computers \& Industrial Engineering}}
  \bibinfo{volume}{161} (\bibinfo{year}{2021}), \bibinfo{pages}{107625}.
\newblock
\showISSN{0360-8352}


\bibitem[Vaswani et~al\mbox{.}(2017)]%
        {NIPS2017_3f5ee243}
\bibfield{author}{\bibinfo{person}{Ashish Vaswani}, \bibinfo{person}{Noam
  Shazeer}, \bibinfo{person}{Niki Parmar}, \bibinfo{person}{Jakob Uszkoreit},
  \bibinfo{person}{Llion Jones}, \bibinfo{person}{Aidan~N Gomez},
  \bibinfo{person}{\L~ukasz Kaiser}, {and} \bibinfo{person}{Illia Polosukhin}.}
  \bibinfo{year}{2017}\natexlab{}.
\newblock \showarticletitle{Attention is All you Need}. In
  \bibinfo{booktitle}{\emph{Advances in Neural Information Processing
  Systems}}, \bibfield{editor}{\bibinfo{person}{I.~Guyon},
  \bibinfo{person}{U.~Von Luxburg}, \bibinfo{person}{S.~Bengio},
  \bibinfo{person}{H.~Wallach}, \bibinfo{person}{R.~Fergus},
  \bibinfo{person}{S.~Vishwanathan}, {and} \bibinfo{person}{R.~Garnett}}
  (Eds.), Vol.~\bibinfo{volume}{30}. \bibinfo{publisher}{Curran Associates,
  Inc.}
\newblock


\bibitem[Wen et~al\mbox{.}(2017)]%
        {Wen2017}
\bibfield{author}{\bibinfo{person}{Ruofeng Wen}, \bibinfo{person}{Kari
  Torkkola}, \bibinfo{person}{Balakrishnan~(Murali) Narayanaswamy}, {and}
  \bibinfo{person}{Dhruv Madeka}.} \bibinfo{year}{2017}\natexlab{}.
\newblock \showarticletitle{A multi-horizon quantile recurrent forecaster}. In
  \bibinfo{booktitle}{\emph{NeurIPS 2017}}.
\newblock


\bibitem[Wu et~al\mbox{.}(2021)]%
        {wu2021autoformer}
\bibfield{author}{\bibinfo{person}{Haixu Wu}, \bibinfo{person}{Jiehui Xu},
  \bibinfo{person}{Jianmin Wang}, {and} \bibinfo{person}{Mingsheng Long}.}
  \bibinfo{year}{2021}\natexlab{}.
\newblock \showarticletitle{Autoformer: Decomposition transformers with
  auto-correlation for long-term series forecasting}.
\newblock \bibinfo{journal}{\emph{Advances in Neural Information Processing
  Systems}}  \bibinfo{volume}{34} (\bibinfo{year}{2021}),
  \bibinfo{pages}{22419--22430}.
\newblock


\bibitem[Wu et~al\mbox{.}(2020)]%
        {wu2020deep}
\bibfield{author}{\bibinfo{person}{Neo Wu}, \bibinfo{person}{Bradley Green},
  \bibinfo{person}{Xue Ben}, {and} \bibinfo{person}{Shawn O'Banion}.}
  \bibinfo{year}{2020}\natexlab{}.
\newblock \showarticletitle{Deep transformer models for time series
  forecasting: The influenza prevalence case}.
\newblock \bibinfo{journal}{\emph{arXiv preprint arXiv:2001.08317}}
  (\bibinfo{year}{2020}).
\newblock


\bibitem[Xu et~al\mbox{.}(2014)]%
        {6559990}
\bibfield{author}{\bibinfo{person}{Yuehai Xu}, \bibinfo{person}{Eitan
  Frachtenberg}, \bibinfo{person}{Song Jiang}, {and} \bibinfo{person}{Mike
  Paleczny}.} \bibinfo{year}{2014}\natexlab{}.
\newblock \showarticletitle{Characterizing Facebook's Memcached Workload}.
\newblock \bibinfo{journal}{\emph{IEEE Internet Computing}}
  \bibinfo{volume}{18}, \bibinfo{number}{2} (\bibinfo{year}{2014}),
  \bibinfo{pages}{41--49}.
\newblock


\bibitem[Yang and Youn(2021)]%
        {9488816}
\bibfield{author}{\bibinfo{person}{Eunju Yang} {and} \bibinfo{person}{Chan-Hyun
  Youn}.} \bibinfo{year}{2021}\natexlab{}.
\newblock \showarticletitle{Individual Load Forecasting for Multi-Customers
  with Distribution-aware Temporal Pooling}. In \bibinfo{booktitle}{\emph{IEEE
  INFOCOM 2021 - IEEE Conference on Computer Communications}}.
  \bibinfo{pages}{1--10}.
\newblock


\bibitem[Yoon et~al\mbox{.}(2022)]%
        {10.1145/3534678.3539348}
\bibfield{author}{\bibinfo{person}{Susik Yoon}, \bibinfo{person}{Youngjun Lee},
  \bibinfo{person}{Jae-Gil Lee}, {and} \bibinfo{person}{Byung~Suk Lee}.}
  \bibinfo{year}{2022}\natexlab{}.
\newblock \showarticletitle{Adaptive Model Pooling for Online Deep Anomaly
  Detection from a Complex Evolving Data Stream}. In
  \bibinfo{booktitle}{\emph{Proceedings of the 28th ACM SIGKDD Conference on
  Knowledge Discovery and Data Mining}} (Washington DC, USA)
  \emph{(\bibinfo{series}{KDD '22})}. \bibinfo{publisher}{Association for
  Computing Machinery}, \bibinfo{address}{New York, NY, USA},
  \bibinfo{pages}{2347–2357}.
\newblock
\showISBNx{9781450393850}


\bibitem[Yu et~al\mbox{.}(2018a)]%
        {yu2018improving}
\bibfield{author}{\bibinfo{person}{Yongjia Yu}, \bibinfo{person}{Vasu Jindal},
  \bibinfo{person}{Farokh Bastani}, \bibinfo{person}{Fang Li}, {and}
  \bibinfo{person}{I-Ling Yen}.} \bibinfo{year}{2018}\natexlab{a}.
\newblock \showarticletitle{Improving the smartness of cloud management via
  machine learning based workload prediction}. In
  \bibinfo{booktitle}{\emph{2018 IEEE 42nd Annual Computer Software and
  Applications Conference (COMPSAC)}}, Vol.~\bibinfo{volume}{2}. IEEE,
  \bibinfo{pages}{38--44}.
\newblock


\bibitem[Yu et~al\mbox{.}(2018b)]%
        {8377827}
\bibfield{author}{\bibinfo{person}{Yongjia Yu}, \bibinfo{person}{Vasu Jindal},
  \bibinfo{person}{Farokh Bastani}, \bibinfo{person}{Fang Li}, {and}
  \bibinfo{person}{I-Ling Yen}.} \bibinfo{year}{2018}\natexlab{b}.
\newblock \showarticletitle{Improving the Smartness of Cloud Management via
  Machine Learning Based Workload Prediction}. In
  \bibinfo{booktitle}{\emph{2018 IEEE 42nd Annual Computer Software and
  Applications Conference (COMPSAC)}}, Vol.~\bibinfo{volume}{02}.
  \bibinfo{pages}{38--44}.
\newblock


\bibitem[Zhou et~al\mbox{.}(2021)]%
        {Zhou2021}
\bibfield{author}{\bibinfo{person}{Haoyi Zhou}, \bibinfo{person}{Shanghang
  Zhang}, \bibinfo{person}{Jieqi Peng}, \bibinfo{person}{Shuai Zhang},
  \bibinfo{person}{Jianxin Li}, \bibinfo{person}{Hui Xiong}, {and}
  \bibinfo{person}{Wancai Zhang}.} \bibinfo{year}{2021}\natexlab{}.
\newblock \showarticletitle{{Informer: Beyond Efficient Transformer for Long
  Sequence Time-Series Forecasting}}.
\newblock \bibinfo{journal}{\emph{35th AAAI Conference on Artificial
  Intelligence, AAAI 2021}}  \bibinfo{volume}{12B} (\bibinfo{year}{2021}),
  \bibinfo{pages}{11106--11115}.
\newblock
\showISBNx{9781713835974}
\showISSN{2159-5399}


\end{thebibliography}





\end{document}